\documentclass[journal]{IEEEtran}

\IEEEoverridecommandlockouts
\usepackage{amsmath,amssymb,amsfonts}
\usepackage{algorithmic}
\usepackage{graphicx}
\graphicspath{{./figures/standalone/}}

\ifCLASSOPTIONcompsoc \usepackage[caption=false,font=normalsize,labelfont=sf,textfont=sf]{subfig}
\else
\usepackage[caption=false,font=footnotesize]{subfig}
\fi

\usepackage{accents}
\usepackage{textcomp}
\def\BibTeX{{\rm B\kern-.05em{\sc i\kern-.025em b}\kern-.08em
    T\kern-.1667em\lower.7ex\hbox{E}\kern-.125emX}}

\usepackage{lipsum} 
 \usepackage{balance}
\usepackage{booktabs} 
\usepackage{enumitem}
\usepackage{mathtools}
\usepackage[lined,boxed,commentsnumbered,linesnumbered,ruled]{algorithm2e}
\usepackage{tikz} 
\usetikzlibrary{shapes, patterns,decorations.text, decorations.pathreplacing}
\usetikzlibrary{external}
\usepackage{soul}

\usepackage[noadjust]{cite} 

\usepackage{pgfplots}
\pgfplotsset{filter discard warning=false}
\pgfplotsset{compat=1.14}
\usepgfplotslibrary{colorbrewer}
\usepgfplotslibrary{groupplots}

\usepackage[capitalize]{cleveref}
\crefname{equation}{\unskip}{\unskip}
\crefname{claim}{Claim}{Claims} 
\usepackage{array}
\usepackage{multirow}
\newcolumntype{C}[1]{>{\centering\arraybackslash}p{#1}}
\allowdisplaybreaks

\graphicspath{{./figures/}}

\newcommand{\E}[1]{\mathbb{E}\left[ #1 \right]}
\newcommand{\Cov}[1]{\operatorname{Cov} \left[ #1 \right]}
\newcommand{\Var}[1]{\operatorname{Var} \left[ #1 \right]}

\newcommand{\ellpd}{\ell_\mathsf{PD}}
\newcommand{\taupd}{\tau_\mathsf{PD}}

\newcommand{\rprocess}{r_1(\taupd)}
\newcommand{\rmean}{\E{\rprocess}}

\newcommand{\estar}{\epsilon^*}
\newcommand{\sstart}{\alpha}
\newcommand{\ssend}{\beta}

\newcommand{\e}{\epsilon}

\newcommand{\m}{\mu_0}
\newcommand{\mb}{\breve{\mu}_0}
\newcommand{\defn}{=}
\newcommand{\der}{\mathrm{d}}
\newcommand{\ti}{\to\infty}
\newcommand{\grows}{\ti}

\newcommand{\FER}{P_\mathsf{f}}
\newcommand{\FERt}{P_\mathsf{f,t}}
\newcommand{\FERu}{P_\mathsf{f,u}}

\newcommand{\doublehat}[1]{\tilde{#1}}

\newcommand{\gammasingle}{\breve{\gamma}}

\newcommand{\sstartsingle}{\breve{\sstart}}
\newcommand{\sstartdouble}{\doublehat{\sstart}}

\newcommand{\ssenddouble}{\doublehat{\ssend}}
\newcommand{\thetasingle}{\breve{\theta}}

\newcommand{\nusingle}{\breve{\nu}}

\newcommand{\dv}{d_\mathsf{v}}
\newcommand{\dc}{d_\mathsf{c}}

\newcommand{\speedpd}{V_{\mathsf{PD}}}
\newcommand{\speedpde}{\speedpd(\e)}
\newcommand{\speedbp}{V_{\mathsf{BP}}}
\newcommand{\speedbpe}{\speedbp(\e)}

\newcommand{\iterstart}{I_{\mathsf{start}}}
\newcommand{\iterend}{I_{\mathsf{end}}}

\newcommand{\iterlim}{I}
\newcommand{\ieff}{I_{\mathsf{eff}}}
\newcommand{\teff}{t_{\mathsf{eff}}}
\newcommand{\tstart}{t_{\mathsf{start}}}
\newcommand{\dmin}{D_{\mathsf{min}}}
\newcommand{\tmin}{\tau_{\mathsf{min}}}
\newcommand{\Leff}{L_{\mathsf{eff}}}

\newcommand{\succdec}{S}
\newcommand{\psuccess}{\operatorname{Pr} \left\{ \succdec \right\}}

\newcommand{\pditer}{n_\mathsf{PD}}

\newcommand{\gammabp}{\gammasingle_{\mathsf{BP}}}
\newcommand{\nubp}{\nusingle_{\mathsf{BP}}}
\newcommand{\thetabp}{\thetasingle_{\mathsf{BP}}}
\newcommand{\vprocessbp}{v_{\mathsf{BP}}}
\newcommand{\vprocessbpi}{v_{\mathsf{BP},i}}
\newcommand{\posprocessbp}{\eta}
\newcommand{\posprocessbpint}{P_\mathsf{L}}

\newcommand{\sigmaiou}{\sigma_2}
\newcommand{\sigmaou}{\sigma_1}
\newcommand{\Wou}{B_\tau}
\newcommand{\Wiou}{B'_\tau}

\newcommand{\Li}{\mathcal{L}}

\newcommand{\oudecay}{b}
\newcommand{\oumean}{m}

\newcommand{\fslope}{c_\mathsf{f}}

\begin{document}

\providecommand{\keywords}[1]
{
    {\small	\textbf{\textit{Index Terms---}#1.}}
}

\title{Finite-Length Scaling of SC-LDPC Codes\\  With a Limited Number of Decoding Iterations}


\author{
    Roman Sokolovskii, \IEEEmembership{Graduate Student Member, IEEE}, Alexandre Graell i Amat, \IEEEmembership{Senior Member, IEEE}, \\and Fredrik Br\"annstr\"om
    \thanks{This work was funded by the Swedish Research Council (grant 2016-4026).}
    \thanks{R. Sokolovskii, A. Graell i Amat, and F. Br\"annstr\"om are with the Communication Systems Group, Department of
    Electrical Engineering, Chalmers University of Technology, SE-41296 Gothenburg, Sweden (email:
    \{roman.sokolovskii,alexandre.graell,fredrik.brannstrom\}@chalmers.se).}}

\maketitle

\begin{abstract}
    We propose four finite-length scaling laws to predict the frame error rate (FER) performance of spatially-coupled low-density parity-check codes under full belief propagation (BP) decoding with a limit on the number of decoding iterations and a scaling law for sliding window decoding, also with limited iterations.
    The laws for full BP decoding provide a choice between accuracy and computational complexity; a good balance between them is achieved by the law that models the number of decoded bits after a certain number of BP iterations by a time-integrated Ornstein-Uhlenbeck process.
    This framework is developed further to model sliding window decoding as a race between the integrated Ornstein-Uhlenbeck process and an absorbing barrier that corresponds to the left boundary of the sliding window.
    The proposed scaling laws yield accurate FER predictions.
\end{abstract}
\keywords{Codes-on-graphs, finite-length code performance, spatially coupled LDPC codes, window decoding}

\section{Introduction}
\label{sec:intro}

Spatially-coupled low-density parity-check (SC-LDPC) \mbox{codes~\cite{ref:Jime99,ref:Lent10}} achieve capacity under suboptimal yet computationally feasible belief propagation (BP) decoding, which was first observed numerically~\cite{ref:Lent10}, then proved for the binary erasure channel (BEC)~\cite{ref:Kude11} and later for the broad class of binary-input memoryless symmetric channels~\cite{ref:Kude13}.
Moreover, it was shown that the minimum distance of regular SC-LDPC code ensembles grows linearly with the block length~\cite{ref:Srid07}.
Spatial coupling is also a powerful technique more broadly; it has been applied to, e.g., turbo-like codes~\cite{ref:Molo17}, product-like codes~\cite{ref:Smit12}, lossy compression~\cite{ref:Aref12}, and compressed sensing~\cite{ref:Dono13}.

The Tanner graph of an SC-LDPC code is constructed by arranging the Tanner graphs of several uncoupled LDPC codes into a sequence and interconnecting them according to a predefined pattern.
The interconnecting is done in such a way so as to create \textit{structured irregularity} at the boundaries of the resulting chain, such that the bits at the boundaries are better protected than those in the middle of the chain and are more likely to be decoded successfully; during BP decoding, information propagates from those boundaries inward in a wave-like fashion.

The structured irregularity at the boundaries entails a loss in code rate.
The rate loss goes to zero as the chain length grows, but so does the probability that the decoding waves fail to propagate through the entire chain.
A longer chain also implies higher latency if BP decoding is performed on the whole received sequence, a scheme we refer to as \emph{full BP}.
To limit decoding latency, so-called \textit{sliding window decoding,} originally proposed in~\cite{ref:Iyen12}, is used in practice.
Sliding window decoding limits BP decoding to a window of several spatial positions that slides through the chain from the left boundary rightward, making a decision on the bits that leave the window along the way and thus limiting decoding latency to the size of the window.
Further, due to constraints on decoding latency and energy efficiency, both full BP and sliding window decoding must in practice be limited in terms of the maximum number of BP iterations.

To adopt an SC-LDPC code in a practical setting, the system designer would have to specify a range of parameters, including the underlying LDPC code ensemble, the interconnecting pattern, the length of the coupled chain, the size of the sliding window, and the limit on the number of decoding iterations.
It is therefore important to understand the influence of these parameters on the error-correcting performance, which is the subject of ongoing scientific inquiry.

\begin{figure*}
    \centering
    \includegraphics[width=.8\textwidth]{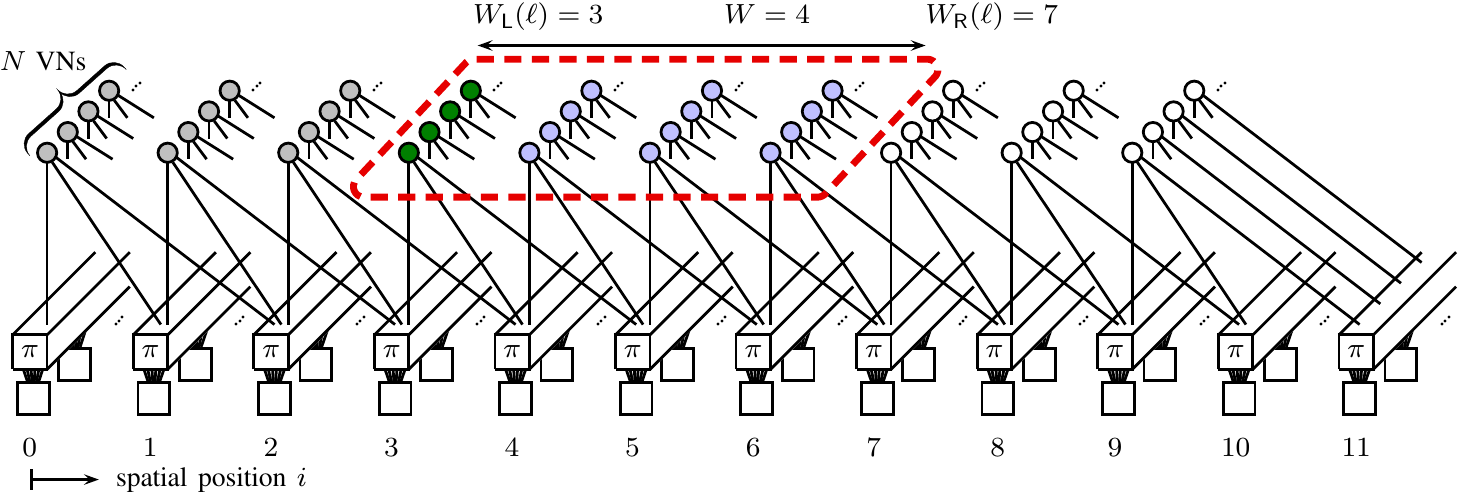}
    \vspace{-3pt}
    \caption{Tanner graph of the terminated $(\dv,\dc,L,N)$ SC-LDPC ensemble with $\dv=3,\dc=6,$
    $L=10$, and $N$ VNs per spatial position. The rectangular cuboids marked with $\pi$ are permutation blocks. The sliding window of length $W=4$ is shown as the red dashed frame. The decision on the VNs in the first $3$ spatial positions (gray) has already been made; the associated values are fixed. The VNs in position $W_\mathsf{L}(\ell) = 3$ (green) are next to be decided upon and fixed. The VNs in the next three positions (blue) participate in message passing, while those in positions $W_\mathsf{R}(\ell) = 7$ and above (white) do not.}
    \label{fig:ensemble}
    \vspace{-2ex}
\end{figure*}

The asymptotic performance of SC-LDPC codes (i.e., when the block length of the component LDPC codes grows large) is relatively well understood.
Much less is known about their finite-length behavior.
Olmos and Urbanke proposed a finite-length scaling law to predict the frame error rate (FER) of SC-LDPC code ensembles under full BP decoding with an unlimited number of iterations over the BEC~\cite{ref:Olmo15}.
The law in~\cite{ref:Olmo15} follows the approach proposed earlier in~\cite{ref:Amra09} for uncoupled LDPC code ensembles and focuses on the number of degree-one check nodes (CNs) available for peeling decoding.
Peeling decoding is equivalent in error-correcting performance to BP decoding with unlimited iterations but is more tractable analytically.
A decoding failure corresponds to the peeling decoder running out of degree-one CNs before recovering the entire codeword.
The authors estimate the probability of that event using an exponential approximation to the first-hit time distribution of an appropriately chosen Ornstein-Uhlenbeck process~\cite{ref:Olmo15}.
This framework was later applied to protograph-based~\cite{ref:Stin16} and generalized~\cite{ref:Cost18} SC-LDPC code ensembles.

The prediction in~\cite{ref:Olmo15} captures the slope of the FER curve well.
However, it exhibits a gap to the simulated FER performance.
In~\cite{ref:Soko20}, we closed this gap by proposing an alternative scaling law that models the number of degree-one CNs available for peeling decoding as a sum of two independent Ornstein-Uhlenbeck processes that correspond to the decoding waves from the left and right boundary of the chain.
We provided laws to also predict the bit and block error rate performance.
Importantly for practical applications, we proposed scaling laws for the frame, bit, and block error rate under sliding window decoding, albeit still in the case of unlimited number of iterations.

The scaling law for the FER under full BP decoding from~\cite{ref:Soko20} was used in~\cite{ref:Kwak21} to elucidate the non-trivial space of trade-offs associated with the choice of code parameters and extended in~\cite{ref:Soko21} to predict the error rate performance of periodically-doped SC-LDPC code ensembles for streaming.
In~\cite{ref:Kwak22}, the authors optimize protograph-based SC-LDPC code constructions under sliding window decoding using a criterion derived from finite-length scaling models (the so-called window mean parameter); they show that taking into account the finite-length scaling behavior of SC-LDPC code ensembles during code optimization can yield codes that significantly outperform their counterparts designed using asymptotic BP thresholds only.

Our ambition is to make finite-length scaling laws the tool of choice in such code parameter optimization.
To that end, this paper tackles the problem of predicting the FER of SC-LDPC code ensembles over the BEC under both full BP and sliding window decoding for the practical case of \textit{a limited number of iterations.}
First, we consider full BP decoding and propose four scaling laws that vary in accuracy and computational complexity.
One of these laws models the number of bits recovered after a certain number of BP iterations by a time-integrated Ornstein-Uhlenbeck process.
We then extend this framework and model sliding window decoding as a race between an absorbing barrier that corresponds to the left boundary of the window and an integrated Ornstein-Uhlenbeck process (with an additional diffusion term) that corresponds to the position of the left decoding wave.
We estimate the probability of the wave being absorbed at the barrier (and thus overtaken by the sliding window) by numerically solving the initial value problem of the corresponding Fokker-Planck equation with appropriately chosen boundary conditions.
The proposed laws yield accurate predictions of the FER and allow us to quantify performance degradation associated with the introduction of the limit on the number of decoding iterations, making practical code parameter optimization using finite-length scaling laws a step closer to reality.

\section{Preliminaries}
\label{sec:preliminaries}

We consider the semi-structured $(\dv,\dc,L,N)$ SC-LDPC code ensemble introduced in~\cite{ref:Olmo15}.
Its Tanner graph is shown in Fig.~\ref{fig:ensemble}.
To construct the Tanner graph of an element of the $(\dv,\dc,L,N)$ SC-LDPC code ensemble, one must first take $L$ Tanner graphs of length-$N$ $(\dv,\dc)$-regular LDPC codes of variable node (VN) degree $\dv$ and CN degree $\dc$ and arrange them into $L$ spatial positions indexed by $i \in \Li=\{0,\ldots,L-1\}$.
Each spatial position contains $N$ VNs and $M\!=\!\frac{\dv}{\dc}N$ CNs, where we assume $M$ is an integer.
We refer to $N$ as the \textit{component code length} and to $L$ as the \textit{chain length}.
The set of all $LN$ VNs in the Tanner graph---and thus of all $LN$ code bits---is referred to as the \textit{frame}.
The $L$ Tanner graphs of the individual (uncoupled) LDPC codes are then interconnected as follows: each VN at position $i \in \Li$ is connected to $\dv$ CNs in positions $[i,\ldots,i + \dv - 1]$.
Specifically, one CN is chosen uniformly at random among $M$ CNs at position $i$, another one at position $i + 1$, and so on until position $i + \dv - 1$.
To connect the overhanging edges at the end of the chain, $\dv - 1$ additional positions that contain CNs only are appended, resulting in a \textit{terminated} ensemble.
The generation of the elements from this ensemble is described in detail in~\cite{ref:Olmo15} and can be expressed in terms of choosing the $L + \dv - 1$ permutation blocks in the Tanner graph in Fig.~\ref{fig:ensemble}, marked by $\pi$.

The ensemble is structured from the VN perspective---it is certain that each VN is connected to CNs in $\dv$ different spatial positions.
The same cannot be said about the CNs.
Indeed, a CN at position $i \in \{ \dv - 1,\ldots,L - 1 \}$ can be connected to $\dc$ VNs from any non-empty subset of positions $[i-\dv+1,\ldots,i]$, depending on how the permutation block at position $i$ reshuffles the edges connected to VNs at positions $[i-\dv+1,\ldots,i]$ (see Fig.~\ref{fig:ensemble}).
This particular ``semi-structured'' ensemble is proposed in~\cite{ref:Olmo15} to simplify the analysis.
The same approach was later applied to protograph-based ensembles~\cite{ref:Stin16}.

In addition to the terminated ensemble, we also consider the \textit{truncated} and \textit{unterminated} ensembles.
In the truncated ensemble, the additional $\dv - 1$ positions with CNs only are not added to the chain of length $L$, and the overhanging edges emanating from VNs at positions $[L-\dv+1,\ldots,L-1]$ are simply deleted from the Tanner graph.
The degree of VNs in these last $\dv - 1$ positions is therefore reduced.
In the unterminated ensemble, the chain is neither terminated nor truncated, resulting in a ``semi-infinite'' sequence of coupled codes.
We introduce the unterminated ensemble for the analysis of sliding window decoding, and we evaluate the decoding error probability over the first $L'$ positions of this semi-infinite chain.

Key to impressive error-correcting performance of SC-LDPC codes under BP decoding is the lower degree of CNs at the terminated boundaries of the chain, i.e., at the left boundary of the truncated and unterminated ensembles and at both left and right boundaries of the terminated ensemble.
During BP decoding, information propagates from the terminated boundaries of the chain inward in a wave-like fashion.
The finite-length scaling laws aim to estimate the probability that such ``decoding waves'' fail to propagate through the entire chain, which results in decoding error.

Full BP decoding entails a decoding latency of $LN$ bits, which is impractical for long spatially-coupled chains.
To limit decoding latency, sliding window decoding is used in practice, where decoding is limited to VNs in a window of $W$ spatial positions (depicted as the red dashed rectangle in Fig.~\ref{fig:ensemble}). 
After a certain number of BP iterations, the decoder decides on the values of the bits in the left-most spatial position within the window (colored green in Fig.~\ref{fig:ensemble}) and the window \textit{slides} by one position to the right over the Tanner graph. 
We index BP decoding iterations by $\ell$ and denote the leftmost position of the window by $W_\mathsf{L}(\ell)$.
The first position just outside the window is denoted by $W_\mathsf{R}(\ell)$, as illustrated in Fig.~\ref{fig:ensemble}.
Sliding window decoding  has a decoding latency of $WN$ bits~\cite{ref:Iyen12}.

This paper investigates the influence of the limit on the number of BP iterations on the error-correcting performance.
In the case of full BP decoding, we denote this limit by $\iterlim$ and treat it is as a parameter of the decoder.
In the case of sliding window decoding, the system designer should choose two parameters instead: the number of BP iterations after which the window slides to position 1, i.e., $W_\mathsf{L}(\ell) = 1$, which we denote by $I_\mathsf{in}$ and whose impact will be clarified later, and, for $W_\mathsf{L}(\ell)\ge 1$, the number of BP iterations before the window slides further, which we denote by $I_\mathsf{s}$.
The total budget of BP iterations in the case of sliding window decoding can be obtained from $I_\mathsf{in}$ and $I_\mathsf{s}$ for a chain of length $L$ as
\begin{equation}
    \iterlim = I_\mathsf{in} + (L - 1) I_\mathsf{s}\,.
    \label{eq:num_iter}
\end{equation}

We consider transmission over the BEC with erasure probability $\e$.

\subsection{Density Evolution}
\label{sse:density_evolution}

Let $q_{i+j,i}^{(\ell)}$ denote the probability that a CN at position $i+j$ sends an erasure message to a VN at position $i$ at BP iteration $\ell$.
Likewise, let $p_i^{(\ell)}$ be the probability that a VN at position $i$ is erased, and $p_{i,i+j}^{(\ell)}$ the probability that a VN at position $i$ sends an erasure message to a CN at position $i+j$ at BP iteration $\ell$.
The CN update for the semi-structured ensemble averages the incoming error probabilities as
\begin{equation}
    q_{i+j, i}^{(\ell)} = 1 - \left( 1 - \frac{1}{\dv}\sum_{j'=0}^{\dv-1}  p_{i+j-j',i+j}^{(\ell-1)} \right)^{\dc - 1}\,.
\label{eq:cnupd_semi}
\end{equation}
To circumvent the problem of the reduced-degree CNs at the boundaries, the values of $p_{i+j-j',i+j}^{(\ell-1)}$ that correspond to VN indices $i+j-j'$ outside the chain---i.e., when $i+j-j' \notin \Li$---are set to zero, implying that the VNs outside the chain are not erased.

Since a VN at position $i\in\Li$ is connected to $\dv$ consecutive positions $\{i, \ldots, i+\dv-1\}$, as we discussed in the beginning of Section~\ref{sec:preliminaries}, the VN update for the semi-structured ensemble is
\begin{equation}
p_{i, i+j}^{(\ell)}=\e \prod_{j^{\prime} \neq j} q_{i+j^{\prime}, i}^{(\ell)} \,,
\label{eq:vnupd}
\end{equation}
and the \textit{a posteriori} probability that a VN at position $i$ remains erased at iteration $\ell$ is
\begin{equation}
    p_{i}^{(\ell)}=\e \prod_{j = 0}^{\dv - 1} q_{i+j, i}^{(\ell)}\,.
\label{eq:vnerased}
\end{equation}

To numerically estimate the BP decoding threshold $\estar$ for a given $(\dv,\dc,L,N)$ SC-LDPC code ensemble, we initialize $p_{i,i+j}^{(0)} = 1$ for all $i \in \Li, j \in \{0,\ldots,\dv-1\}$ and iterate equations~\eqref{eq:cnupd_semi}--\eqref{eq:vnupd} until the VN erasure probability~\eqref{eq:vnerased} converges either to zero (for $\e \le \estar$) or to another fixed point (for $\e > \estar$) for all $i$.

\subsection{Peeling Decoding}
\label{sse:peeling_decoding}

The finite-length analysis of BP decoding for the BEC becomes more tractable by considering \textit{peeling decoding}~\cite{ref:Luby97}, which is equivalent to BP decoding for an infinite number of iterations.
The peeling decoder gets stuck in the same stopping sets and therefore yields the same performance as the BP decoder~\cite{ref:Luby97}.
At the initial stage of peeling decoding, all VNs that correspond to non-erased bits are removed from the Tanner graph.
At every subsequent iteration, the decoder selects one degree-one CN uniformly at random among all degree-one CNs in the graph.
Since the value of the code bit associated with the VN connected to the chosen degree-one CN can be recovered, the decoder removes both the CN and VN from the Tanner graph along with $\dv$ adjacent edges and modifies the parity-check equations associated with the adjacent CNs according to the value of the recovered bit.
This may in turn create new degree-one CNs to choose from (or remove some other degree-one CNs collaterally).
Each iteration of peeling decoding produces a new \textit{residual} graph, indexed by iteration number $\ellpd$.
Decoding is successful if eventually the decoder manages to peel off all VNs from the original Tanner graph, resulting in an empty graph.
This happens if at every iteration of peeling decoding there is at least one degree-one CN to choose from.
On the other hand, if the decoder runs out of degree-one CNs before recovering all VNs, decoding gets trapped in a stopping set and fails.

The goal of scaling laws for LDPC codes in~\cite{ref:Amra09} and for SC-LDPC codes in~\cite{ref:Olmo15, ref:Soko20} is to estimate the error rate in the waterfall region, which is dominated by the stopping sets linearly sized with $N$~\cite{ref:Amra09}.
A scaling law therefore aims to estimate the probability that the number of VNs remaining in the residual graph when decoding stops is linearly sized with $N$.



\subsection{The Scaling Laws for Unlimited Number of Iterations}
\label{sse:old_laws}

The general approach to finite-length scaling of SC-LDPC code ensembles originally proposed in~\cite{ref:Olmo15} for full BP and improved and extended to sliding-window decoding  in~\cite{ref:Soko20} is to focus on the stochastic process associated with the number of degree-one CNs in the residual graphs during peeling decoding  normalized by $N$~\cite{ref:Luby01},
\begin{equation}
    \rprocess \defn \frac{1}{N}\sum_u R_{1,u}(\taupd)\,,
    \label{eq:rprocess}
\end{equation}
where $\taupd \defn \ellpd / N$ is the normalized time of peeling decoding, and $R_{1,u}(\taupd)$ is the number of degree-one CNs at position $u$ at iteration~$\ellpd$.
Since peeling decoding requires at least one degree-one CN at every iteration, a decoding error occurs if $\rprocess$ hits zero before recovering all VNs that are erased by the channel.
We refer to a realization of $\rprocess$ as a \textit{decoding trajectory}.
\begin{figure}[!t]
    \centering
    \includegraphics[width=\columnwidth]{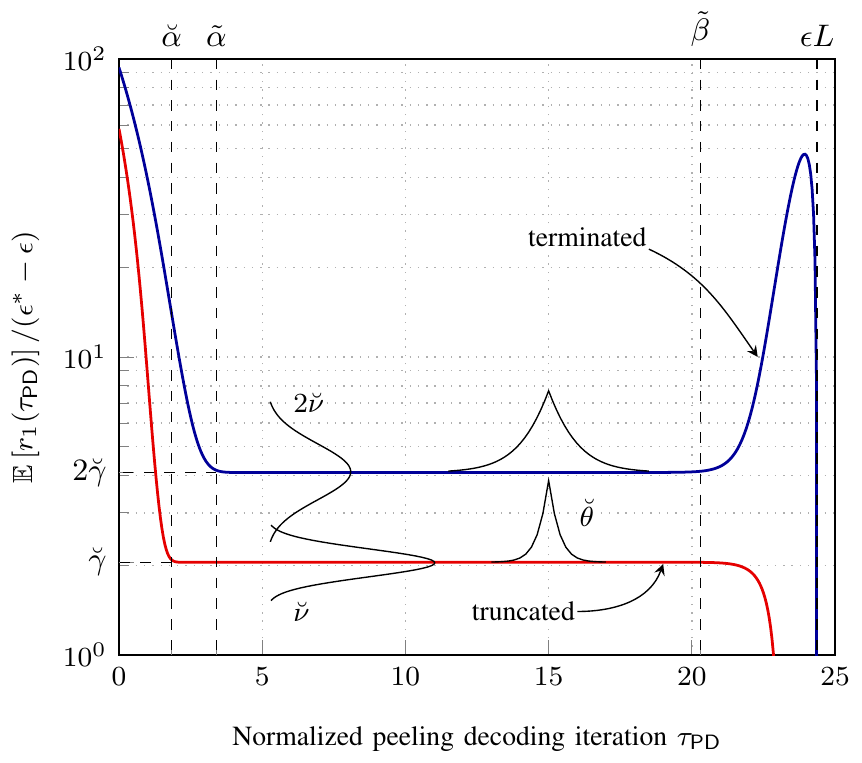}
    \vspace{-18pt}
    \caption{The evolution of $\rmean$ during peeling decoding, normalized by the distance to the BP threshold, for the $(5,10,L\!=\!50)$ ensemble with $\estar\!=\!0.4994$ at $\e\!=\!0.4875$.}
    \label{fig:parameters}
    \vspace{-4pt}
\end{figure}

For a fixed $\taupd$, the distribution of $\rprocess$ converges to a Gaussian as $N\grows$ and concentrates around its mean $\rmean$ with expectation taken over the ensemble, channel, and peeling decoding realizations~\cite{ref:Amra09,ref:Olmo15}.
The evolution of $\rmean$ over $\taupd$ can be obtained by numerically solving a system of coupled differential equations called \textit{mean evolution}.
Mean evolution equations for the semi-structured $(\dv,\dc,L,N)$ SC-LDPC code ensemble are provided in~\cite{ref:Olmo15}.
For illustration, Fig.~\ref{fig:parameters} shows the mean evolution curves $\rmean$ normalized by $\epsilon^*-\epsilon$ for the terminated (blue curve) and truncated (red curve) $(5,10,L\!=\!50,N)$ SC-LDPC code ensemble at $\e\!=\!0.4875$.

Notably, $\rmean$ exhibits a steady-state phase where it remains essentially constant~\cite{ref:Olmo15}.
We denote the range of $\taupd$ corresponding to the steady state of the terminated ensemble as~$\left[\sstartdouble, \ssenddouble\right]$.
Here and in the following, we denote the variables associated with the terminated ensemble with a tilde, e.g., $\sstartdouble$, and those associated with the truncated ensemble with a breve, e.g., $\sstartsingle$---the two ``extrema'' in a tilde allude to the presence of two decoding waves in the terminated ensemble, and a single ``extremum'' in a breve to a single wave in the truncated ensemble.

During the steady state, the two waves propagating in the terminated chain are each equivalent to the single wave present in the truncated chain.
In~\cite{ref:Soko20}, we proposed to isolate a single wave by focusing on the truncated chain and modeling the first two moments of $\rprocess$ in the same way as is done for the terminated chain in~\cite{ref:Olmo15}, namely
\begin{align}
    \rmean &\approx \gammasingle \left( \estar - \e \right)\,, \label{eq:rprocess_mean} \\
    \Var{ \rprocess } &\approx \frac{\nusingle}{N}\,, \label{eq:rprocess_var} \\
    \Cov{ r_1\!\left(\taupd^{(0)}\right), r_1\!\left(\taupd^{(1)}\right) } &\approx \frac{\nusingle}{N} \exp\!\left(\!-\thetasingle \left| \taupd^{(0)}\!-\!\taupd^{(1)} \right| \right)\,, \label{eq:rprocess_cov}
\end{align}
for a triple of real positive numbers $(\gammasingle,\nusingle,\thetasingle)$.
\begin{figure}[!t]
    \centering
    \includegraphics[width=\columnwidth]{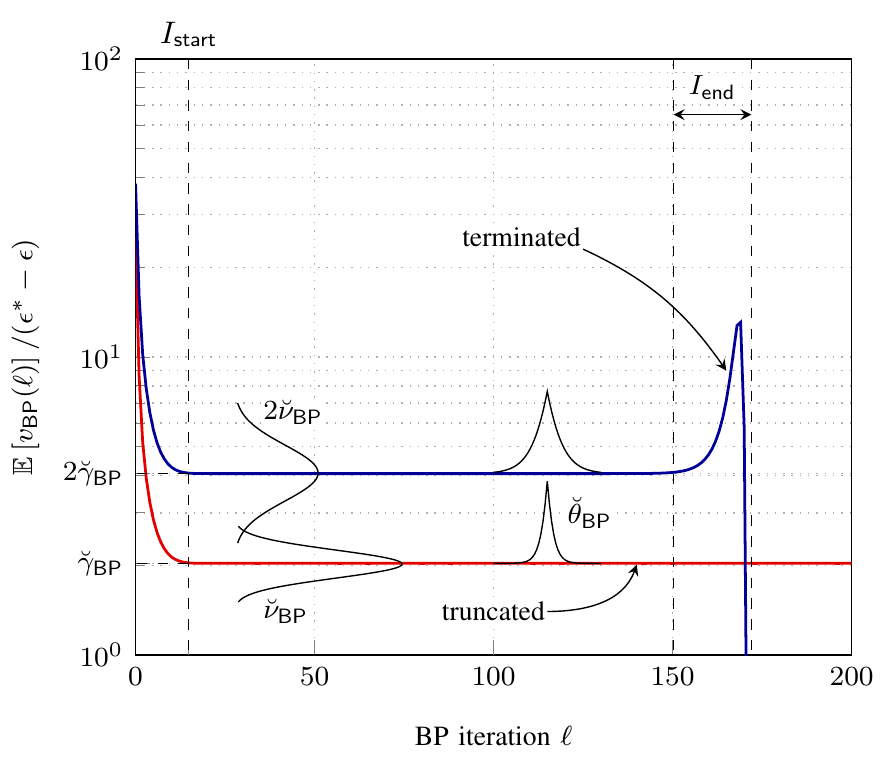}
    \vspace{-18pt}
    \caption{The evolution of $\E{\vprocessbp(\ell)}$ during BP decoding, normalized by the distance to the BP threshold, for the $(5,10,L\!=\!50)$ ensemble with $\estar\!=\!0.4994$ at $\e\!=\!0.47$.}
    \label{fig:de_illustration}
    \vspace{-4pt}
\end{figure}

Apart from the BP decoding threshold $\estar$, the scaling law for unlimited number of BP iterations requires the five parameters $(\sstartdouble,\ssenddouble,\gammasingle,\nusingle,\thetasingle)$.
The first three can be estimated by numerically solving mean evolution equations.
We estimate them for several $\e$ and linearly interpolate the values in between.
The last two can be estimated by solving an augmented system of differential equations called~\textit{covariance evolution}~\cite{ref:Olmo15}.
Instead, we estimate $(\nusingle,\thetasingle)$ from a set of realizations of $\rprocess$ for a fixed $(\e,N)$ and treat them as parameters that depend on $(\dv,\dc)$ only.
The meaning of the scaling parameters for peeling decoding is illustrated in Fig.~\ref{fig:parameters}.

The key idea behind the scaling laws in~\cite{ref:Olmo15,ref:Soko20} is to model $\rprocess$ in the steady state by an Ornstein-Uhlenbeck process, parametrized to match the first two moments of $\rprocess$~\eqref{eq:rprocess_mean}--\eqref{eq:rprocess_cov}.
Olmos and Urbanke model the steady state of the terminated ensemble by a single Ornstein-Uhlenbeck process~\cite{ref:Olmo15}; we proposed instead a refined model that relies on two independent Ornstein-Uhlenbeck processes and yields more accurate predictions~\cite{ref:Soko20}.
We summarize the model in~\cite{ref:Soko20} here and use it as the starting point for our analysis of decoding with a limited number of iterations.

To estimate the FER, we consider the normalized time of peeling decoding at which the number of degree-one CNs\textemdash and hence the value of $\rprocess$\textemdash drops to zero, referred to as the first-hit time $\tau_0$,
\begin{equation}
    \tau_0 \defn \min \{\taupd : \rprocess = 0 \}\,.
\end{equation}

For the terminated ensemble, to exhaust all available degree-one CNs, the peeling decoder must run out of them in both the left and right decoding wave.
Accordingly, we model $\tau_0$ as the sum of two independent variables $A$ and $B$ that correspond to the first-hit time of the left and right wave, respectively~\cite{ref:Soko20}, as
\begin{equation}
    \tau_0 = A + B\,.
\end{equation}

The number of degree-one CNs available to each wave is modeled as an independent Ornstein-Uhlenbeck process.
It is known that the first-hit time distribution for an Ornstein-Uhlenbeck process converges to an exponential distribution with mean $\mb$ as $N \grows$~\cite{ref:Olmo15},
\begin{equation}
    A, B \sim \operatorname{Exp}(\mb)\,,
    \label{eq:expon_dist_approx}
\end{equation}
where $\mb=\m(\gammasingle,\nusingle,\thetasingle)$ and
\begin{equation}
    \m(\gamma,\nu,\theta) = \frac{\sqrt{2\pi}}{\theta}\int_0^{\gamma\sqrt{N/\nu}\left( \estar - \e \right)}
    \Phi(z)\mathrm{e}^{\frac{1}{2}z^2}\der z\,.
    \label{eq:olmos_mean}
\end{equation}
The PDF of the number of degree-one CNs recovered by the left and right wave during the steady state is
\begin{equation}
    f_A(x) = f_B(x) \approx \mb^{-1} \exp\left(-\frac{x}{\mb}\right)\,.
    \label{eq:expon_pdf_approx}
\end{equation}

Let $\succdec$ denote successful decoding.
Since the total number of degree-one CNs that must be recovered during the steady state of the terminated ensemble is $\ssenddouble - \sstartdouble$, the probability of successful decoding can be expressed as
\begin{align}
    \psuccess &= \operatorname{Pr}\left\{ A + B > \ssenddouble - \sstartdouble\right\} \label{eq:psuccess_unlim} \\
    &= \int\limits_{\ssenddouble - \sstartdouble}^{\infty} \int\limits_{0}^{x} f_A (z) f_B (x-z) \der z \der x \nonumber\,.
\end{align}

The approximation for the FER in~\cite{ref:Soko20} is obtained from~\eqref{eq:psuccess_unlim} and~\eqref{eq:expon_pdf_approx} as
\begin{equation}
    \FERt^{(L)} \approx 1 - \left( 1 + \frac{\ssenddouble - \sstartdouble}{\mb} \right)
    \exp \left( - \frac{\ssenddouble - \sstartdouble}{\mb} \right)\,.
    \label{eq:our_fer}
\end{equation}

In a similar manner, the FER of an unterminated SC-LDPC code ensemble evaluated over $L'$ spatial positions is approximated in~\cite{ref:Soko20} as
\begin{align}
\FERu^{(L')} \approx 1 - \exp\left(-\frac{\epsilon L' - \sstartsingle}{\mb}\right)\,,
\label{eq:untF}
\end{align}
which is the same approximation used by Olmos and Urbanke for the \textit{terminated} ensemble, but with the scaling parameters estimated from the truncated ensemble instead of from the terminated ensemble.

The scaling law for the unterminated ensemble is used here and in~\cite{ref:Soko20} in the analysis of sliding window decoding, where the sliding window does not allow the right wave to propagate to the left by further than $W$ positions, effectively limiting the first $L-W$ positions to decoding by a single wave only.
As we showed in~\cite{ref:Soko20}, this results in a ``two-phase'' decoding, where the first phase comprises the first $L-W$ positions and includes a single decoding wave from the left, and the second phase comprises the last $W$ positions and may contain both the wave from the left and the wave from the right. The FER in the case of sliding window decoding with unlimited number of BP iterations can then be approximated as~\cite{ref:Soko20}
\begin{align}
    P_{\mathsf{f,t,sw}}^{(L,W)} = 1-\left(1-P_{\mathsf{f,u}}^{(L-W)}\right)\left(1-P_{\mathsf{f,t}}^{(W)}\right)\,,
    \label{eq:PfSWb}
\end{align}
where $P_{\mathsf{f,t}}^{(\cdot)}$ is given in \eqref{eq:our_fer}, and $P_{\mathsf{f,u}}^{(\cdot)}$ is given in \eqref{eq:untF}.

Lastly, we will use the \textit{speed} of a decoding wave in our analysis.
Specifically, we assume that a wave traverses $\speedpd$ positions in $N$ peeling decoding iterations.
We estimate $\speedpd$ from the average number of erased VNs in the middle of the coupled chain during the steady state as
\begin{equation}
    \speedpd \approx N\cdot \E{V_{\lfloor L/2 \rfloor} \left(\frac{\ssenddouble - \sstartdouble}{2}\right)}^{-1}\,,
    \label{eq:calc_speed}
\end{equation}
where $\E{V_u(\taupd)}$, the average number of erased VNs at position $u$ at normalized iteration $\taupd$, is produced alongside
$\rmean$ by numerically solving mean evolution~\cite{ref:Soko20}.

\section{The Speed of the Decoding Waves and the Duration of the Steady State}
\label{sec:speed}

Besides providing BP thresholds for SC-LDPC code ensembles, density evolution, described in Section~\ref{sse:density_evolution}, can be used to answer the following questions: How many BP iterations are needed before the steady state begins and the decoding waves establish? How fast do decoding waves propagate under BP decoding? How many iterations does it take for the decoding waves to collapse once they meet at the end of the steady state?

Let $\vprocessbpi(\ell)$ be the fraction of code bits at position $i$ recovered in BP iteration $\ell$.
The expected value of $\vprocessbpi(\ell)$ can be obtained from the decrease in the erasure probability across iterations of density evolution as
\begin{equation}
    \E{\vprocessbpi(\ell)} =  p_i^{(\ell-1)} - p_i^{(\ell)} 
   \end{equation}
with $p_i^{(\ell)}$ from~\eqref{eq:vnerased}.
Define $\vprocessbp(\ell)\triangleq \sum_{i \in \Li}\vprocessbpi(\ell)$.
The number of code bits recovered in iteration $\ell$ is then $N\!\cdot \vprocessbp(\ell)$, with $\E{N\!\cdot \vprocessbp(\ell)}=N\cdot\E{ \vprocessbp(\ell)}$, where
\begin{equation}
    \E{\vprocessbp(\ell)} = \sum_{i \in \Li}\E{ \vprocessbpi(\ell)}\,.
    \label{eq:vbp}
\end{equation}

The evolution of $\E{\vprocessbp(\ell)} / \left( \estar - \e\right)$ through BP iterations is shown in Fig.~\ref{fig:de_illustration} for the terminated (blue curve) and truncated (red curve)  $(5,10,L=50,N)$ SC-LDPC code ensemble at $\e = 0.47$.
In contrast to the mean evolution curves (Fig.~\ref{fig:parameters}), the onset of the steady state happens at the same time for the truncated and terminated ensemble because BP decoding resolves all code bits corresponding to VNs connected to degree-one CNs in parallel.

Denote the speed of the decoding waves under BP decoding by $\speedbp$, measured in positions traveled by the wave per BP iteration, the number of BP iterations before the onset of the decoding waves by $\iterstart$, and the number of BP iterations it takes for the waves to collapse once they meet by $\iterend$.
The parameters $\iterstart$ and $\iterend$ are illustrated in Fig.~\ref{fig:de_illustration}.
We estimate $\speedbp$ by tracking the mid-point of the wave fronts of $p_i^{(\ell)}$ over density evolution iterations, and $(\iterstart, \iterend)$ by tracking the second difference of $\E{\vprocessbp(\ell)}$ across iterations and comparing it with a numerical threshold (the value of $10^{-2}$ is used; the average change in the erasure probability is normalized by the distance to the BP threshold prior to the comparison with the threshold).
All these parameters are estimated for several values of $\e$.
Linear interpolation is used to estimate the intermediate values.
A similar procedure was used in~\cite{ref:Soko20} to estimate $(\gammasingle, \sstartsingle, \sstartdouble, \ssenddouble)$ for peeling decoding.

As introduced in Section~\ref{sec:preliminaries}, we denote the limit on the total number of BP iterations by $\iterlim$.
For decoding to be successful, the decoding waves must meet before the \emph{effective deadline} of
\begin{equation}
    \ieff \defn \iterlim - \iterstart - \iterend
    \label{eq:iterlim_eff}
\end{equation}
iterations of BP decoding after the beginning of the steady state phase.

\begin{figure}[!t]
    \centering
    \includegraphics[width=\columnwidth]{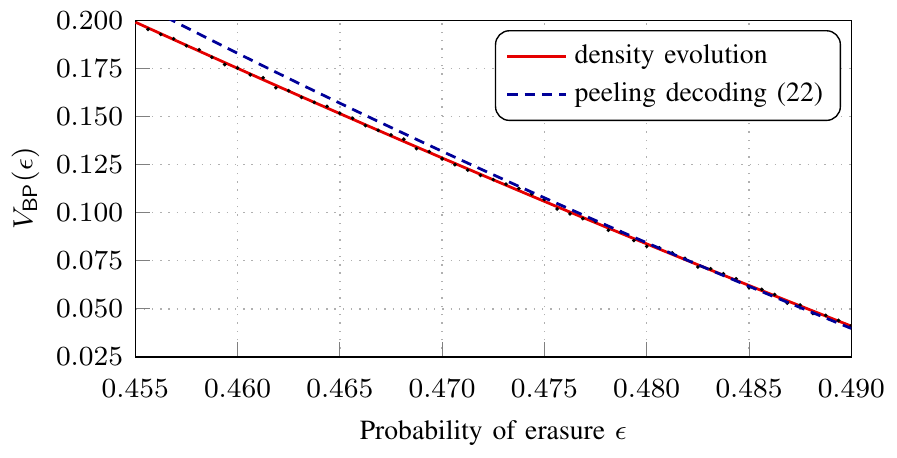}
    \vspace{-18pt}
\caption{The speed of the waves for the $(5,10,L)$ SC-LDPC code ensemble under BP decoding.}
\label{fig:speed_waves}
    \vspace{-2ex}
\end{figure}

Fig.~\ref{fig:speed_waves} shows the speed of the decoding waves during the steady state of BP decoding estimated from density evolution as a function of $\e$, $\speedbpe$.
The black dots represent the values estimated directly, and the red solid line is a quadratic fit to these data.

There is an alternative way to estimate $\speedbpe$.
Since a single iteration of BP decoding recovers all VNs connected to degree-one CNs, we assume that a single BP iteration during the steady state is equivalent to as many peeling decoding iterations as there are degree-one CNs available for each of the waves.
This assumption allows us to approximate $\speedbpe$ as
\begin{equation}
    \speedbpe \approx \speedpde \cdot \gammasingle (\estar - \e)\,,
    \label{eq:speed_conversion}
\end{equation}
where $\speedpde$ is the speed of the waves under \textit{peeling decoding}, which can be calculated as shown in~\eqref{eq:calc_speed}, and  $\gammasingle (\estar - \e)$ is  the average number of degree-one CNs available to one decoding wave during the steady state of peeling decoding (see~\eqref{eq:rprocess_mean}).

The approximation~\eqref{eq:speed_conversion} is also shown in Fig.~\ref{fig:speed_waves} (blue dashed).
The good match between the direct density evolution-based estimation of $\speedbpe$ (red) and the approximation in~\eqref{eq:speed_conversion} (blue dashed) indicates that the number of degree-one CNs available to a decoding wave can be used as a crucial bridge between the steady-state behavior of peeling decoding and BP decoding.

\section{Finite-Length Scaling: Constant\\ Propagation Model}
\label{sec:fl_constant_speed}

For a limited number of BP iterations, the probability of successful decoding~\eqref{eq:psuccess_unlim} should be rewritten as
\begin{equation}
    \psuccess=\operatorname{Pr}\left\{A\!+\!B>\ssenddouble\!-\!\sstartdouble ~ \bigcap \text{~enough iterations} \right\}\,.
    \label{eq:psuccess}
\end{equation}

Density evolution shows that the decoding waves propagate throughout the chain with a constant speed in the limit of $N \grows$.
A natural first step toward a finite-length scaling law would be to also assume a constant propagation speed in the context of a finite $N$---i.e., to assume that with every BP iteration each wave travels by exactly $\speedbp$ positions.

To understand why a limit on the number of iterations affects the probability of successful decoding, consider the following scenario: imagine one of the waves fails immediately at the beginning of the chain; imagine further that $\speedbp$ is too small for the other wave to propagate through the whole chain and meet the first wave within $\ieff$ iterations.
In that case decoding would fail even though it could have been successful without a limit on $\iterlim$.

\begin{figure}[!t]
    \centering
    \includegraphics[width=252pt]{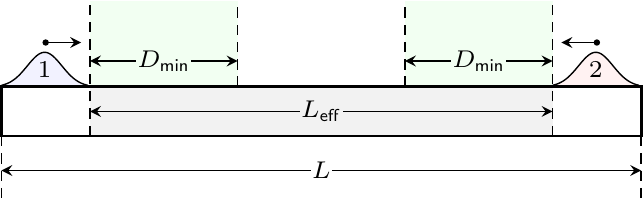}
    \vspace{-18pt}
    \caption{Schematic illustration of the model.}
\label{fig:notation_illustration}
    \vspace{-2ex}
\end{figure}

Our approach is to incorporate a limit on the number of BP iterations into the scaling laws for \textit{peeling decoding}---analyzing BP decoding directly proves to be challenging.
First, we need to account for the width of the decoding waves under peeling decoding. It takes $\ssenddouble - \sstartdouble$ normalized peeling decoding iterations for the waves to propagate through the chain with a constant speed of $\speedpd$ positions per normalized peeling decoding iteration (see Fig.~\ref{fig:parameters}).
This means that the length (in positions) covered during the steady state is
\begin{equation}
    \Leff \defn (\ssenddouble - \sstartdouble) \speedpd\,.
    \label{eq:leff}
\end{equation}
We schematically illustrate the meaning of $\Leff$, which can be thought of as the effective length of the chain, in Fig.~\ref{fig:notation_illustration}.

How much of that effective length should a wave cover under a limit on the number of BP iterations?
Without loss of generality, take the left decoding wave, marked by $1$ in Fig.~\ref{fig:notation_illustration}.
In $\ieff$ BP iterations during the steady state, the wave can propagate by up to $\speedbp \ieff$ positions.
This means that for successful decoding, wave $2$ should propagate by at least
\begin{equation}
    \dmin = \max \left\{ 0, \Leff - \speedbp \ieff \right\}
    \label{eq:dmin}
\end{equation}
positions to meet wave $1$ before the deadline, as illustrated in Fig.~\ref{fig:notation_illustration}.
We need to bear in mind that, \textit{mutatis mutandis}, the same logic applies to both the first and second wave, hence both waves need to propagate by at least $\dmin$ positions for decoding to be successful.

\begin{figure}[!t]
    \centering
    \includegraphics[width=\columnwidth]{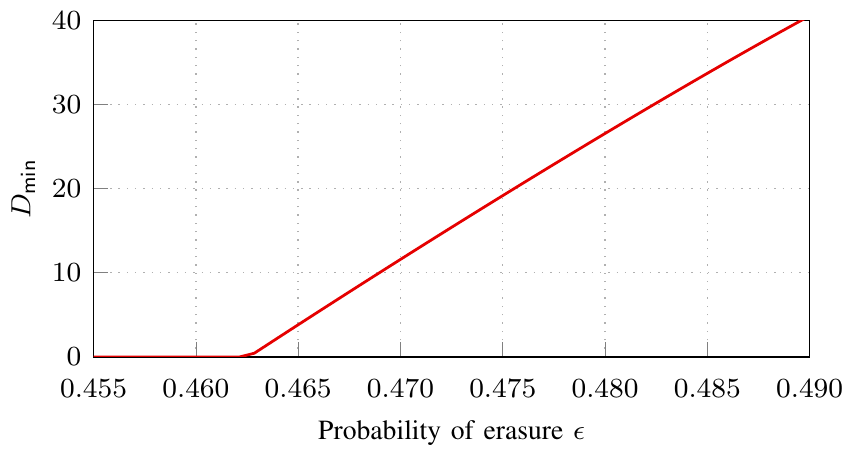}
    \vspace{-18pt}
    \caption{The minimum propagation distance $\dmin$~\eqref{eq:dmin} for the $(5,10,L=50)$ SC-LDPC code ensemble under BP decoding for $\iterlim = 350$.}
\label{fig:dmin}
    \vspace{-2ex}
\end{figure}

Fig.~\ref{fig:dmin} shows an example of $\dmin$ calculated using~\eqref{eq:dmin}.
We observe that $\dmin$ is zero up to a certain value of $\e$ and then increases with increasing $\e$.
$\dmin = 0$ means that a single decoding wave has enough time to propagate all the way along the chain (i.e., successful decoding is possible even if the other wave fails immediately after the onset of the steady state), so the limit on the number of iterations does not change the probability of decoding error compared to the case of unlimited number of iterations.
For large $\e$, $\dmin$ may exceed $\Leff / 2$, which is when the waves do not get a chance to meet in time at all and decoding is bound to fail.

We can now use $\dmin$ to translate the limitation on the number of BP iterations into the language of the first-hit times $A$ and $B$.
To require a wave to travel by at least $\dmin$ positions is to ask it to survive for at least $\tmin \defn \dmin / \speedpd$ peeling decoding iterations.
Bearing in mind that \textit{both} waves need to survive that long to meet the deadline, we can rewrite the second condition in~\eqref{eq:psuccess} as
\begin{equation}
    \text{enough iterations} ~ \Longleftrightarrow ~ A,B > \tmin\,.
\end{equation}
Since the first-hit times $A$ and $B$ are exponentially distributed~\eqref{eq:expon_dist_approx}, the two conditions in~\eqref{eq:psuccess} can be combined as
\begin{align}
    \psuccess &= \operatorname{Pr}\left\{ A + B > \ssenddouble - \sstartdouble ~\bigcap ~ A,B > \tmin \right\} \label{eq:psuccess_speed} \\
    &= \int\limits_{\ssenddouble - \sstartdouble}^{\infty} \int\limits_{\tmin}^{x - \tmin} f_A (z) f_B (x-z) \der z \der x \nonumber \\
    &= \left( 1 + \frac{\ssenddouble - \sstartdouble - 2\tmin}{\mb} \right) \exp \left( - \frac{\ssenddouble - \sstartdouble}{\mb} \right) \nonumber\,,
\end{align}
where we assumed $\dmin < \Leff / 2$; otherwise, $\psuccess = 0$.
\begin{figure}[!t]
    \centering
    \includegraphics[width=\columnwidth]{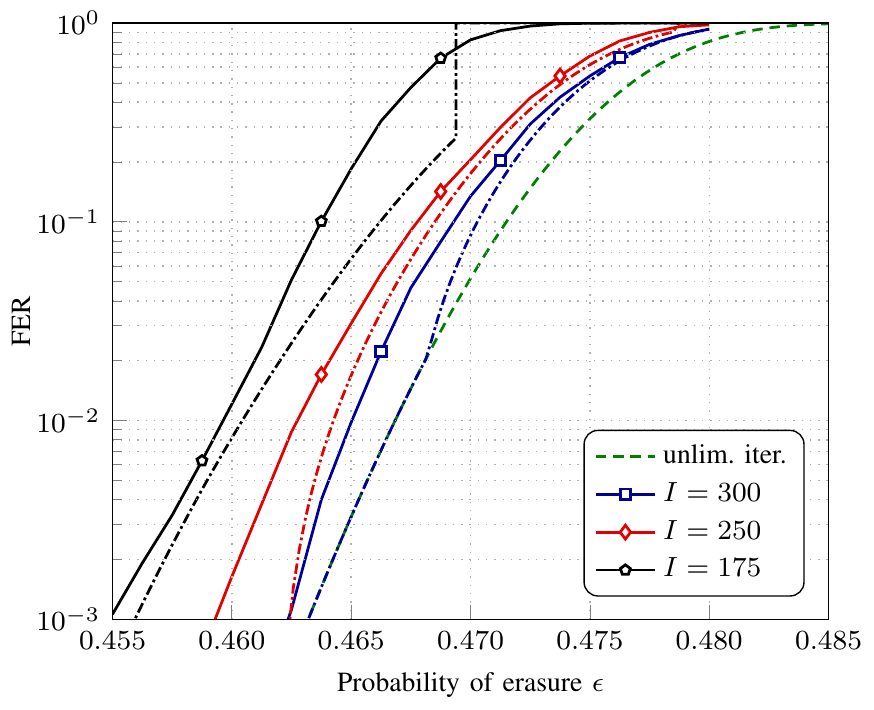}
    \vspace{-18pt}
    \caption{FER for the $(5,10,L=50,N=1000)$ SC-LDPC code ensemble under BP decoding for different limits on the number of iterations $\iterlim$ (solid curves) and its approximation using~\eqref{eq:fl_speed} (corresponding dash-dotted curves).}
\label{fig:fl_speed}
    \vspace{-2ex}
\end{figure}

The FER can therefore be estimated as
\begin{equation}
    \FER \approx 1 - \left( 1 + \frac{\ssenddouble - \sstartdouble - 2\tmin}{\mb} \right) \exp \left( - \frac{\ssenddouble - \sstartdouble}{\mb} \right)
    \label{eq:fl_speed}
\end{equation}
if $\dmin < \Leff / 2$ and $1$ otherwise.

Fig.~\ref{fig:fl_speed} compares the simulated FERs (solid curves) for the terminated $(5,10,L=50,N=1000)$ SC-LDPC code ensemble with the approximation~\eqref{eq:fl_speed} (dash-dotted curves) under different limits on the number of BP iterations $\iterlim = \{ 175, 250, 300 \}$.
Naturally, as the limit $\iterlim$ increases, the solid curves approach the FER curve for the unlimited iterations (green dashed) calculated using~\eqref{eq:our_fer}.
The scaling law~\eqref{eq:fl_speed} is a reasonably accurate approximation to the simulated FER, especially given that the only change to the scaling law for unlimited iterations~\eqref{eq:our_fer} is the introduction of the additional term $-2\tmin$ in~\eqref{eq:fl_speed}.
However, the analytical approximation does not capture the simulated FER behavior in the regions where $\dmin$ is either close to zero (where the approximation curve joins the green dashed curve, as for $\iterlim=300$ at around $\e = 0.468$) or to $\Leff / 2$ (where there may be a discontinuous jump to $\FER = 1$, as for $\iterlim=175$ at around $\e = 0.47$).

\section{Finite-Length Scaling: Randomized Propagation Distance Models}
\label{sec:fl_pditer}

In this section, we introduce three scaling laws that drop the assumption that a decoding wave propagates by the same distance in every BP iteration.
Indeed, the partial mismatch of the scaling law~\eqref{eq:fl_speed} to the simulated FER curves in Fig.~\ref{fig:fl_speed} suggests that the assumption of constant wave propagation does not hold in practice for a finite number of iterations---when $\speedbpe$ is sufficient for the limitation on the number of iterations not to matter for the scaling law (i.e., when predicted FER curves merge with that for full BP decoding), the waves may still fail to meet before the deadline.
Likewise, when $\speedbpe$ becomes so low that the law predicts the FER to be equal to one, the waves may still succeed.

Let us denote by $\pditer(K)$ the number of normalized peeling decoding iterations (and hence the number of VNs recovered) that corresponds to $K$ BP iterations for a decoding wave that has not run out of degree-one CNs.
The scaling law in Section~\ref{sec:fl_constant_speed} and the conversion between $\speedpd$ and $\speedbp$ in~\eqref{eq:speed_conversion} effectively assume $\pditer(K)$ to be constant and equal to $\gammasingle (\estar - \e) K$.
In this section, we treat $\pditer(K)$ as a random variable instead.

A decoding wave stops either because it runs out of degree-one CNs or because it reaches the limit on the number of BP iterations, whichever event happens first.
Correspondingly, let $X_1$ and $X_2$ denote the number of VNs recovered by the first and second wave by that time,
\begin{equation}
    \begin{aligned}
        X_1 &\defn \min\left\{ A, \pditer^{(1)}(\ieff) \right\}\,,\\
        X_2 &\defn \min\left\{ B, \pditer^{(2)}(\ieff) \right\}\,,\\
    \end{aligned}
    \label{eq:numrecovered}
\end{equation}
where $A$ and $B$ are the first-hit times of the first and second wave, respectively,
and the superscripts to $\pditer(\ieff)$ emphasize that these are two independent random variables that correspond to the two decoding waves.
(The two waves are numbered left to right as shown in Fig.~\ref{fig:notation_illustration}.)
The probability of a successful decoding can then be rewritten as
\begin{align}
    \psuccess &= \operatorname{Pr}\left\{ X_1 + X_2 > \ssenddouble - \sstartdouble \right\}\,. \label{eq:psuccess_dist}
\end{align}
Essentially, the model~\eqref{eq:psuccess_dist} incorporates the second requirement in~\eqref{eq:psuccess}---that there should be enough BP iterations for the two waves to meet---into $X_1$ and $X_2$ via a random variable $\pditer(\ieff)$, as opposed to incorporating it via a constant boundary $\tmin$ as in~\eqref{eq:psuccess_speed}.

Assuming $A$ and $\pditer(\ieff)$ to be independent and $A$ to be exponentially distributed (see~\eqref{eq:expon_dist_approx}), we approximate the PDF of $X_1$ as
\begin{align}
    f_X(x) &= f_A(x) \Big[1 - F_{\pditer}(x) \Big] + f_{\pditer}(x) \Big[ 1 - F_A(x) \Big] \nonumber \\
    &\approx\!\frac{1}{\mb} \exp\left( -\frac{x}{\mb} \right) \Big[1\!-\!F_{\pditer}(x) \Big]  \nonumber \\
    &\phantom{=} + f_{\pditer}(x) \exp\left( - \frac{x}{\mb} \right)\,,\label{eq:pdf_x_approx}
\end{align}
where $f_{\pditer}$ and $F_{\pditer}$ denote the PDF and the CDF of $\pditer(\ieff)$, respectively.
The same logic applies to the second wave, so the PDF of $X_2$ is also $f_X(x)$.
We can now use~\eqref{eq:pdf_x_approx} to estimate the FER similarly to~\eqref{eq:psuccess_unlim} as
\begin{align}
    \FER &= 1 - \psuccess \approx 1 - \int\limits_{\ssenddouble - \sstartdouble}^{\infty} \int\limits_0^x f_X(z)f_X(x - z) \der z \der x\,.\label{eq:fer_min}
\end{align}

\begin{figure}[!t]
    \centering
    \includegraphics[width=\columnwidth]{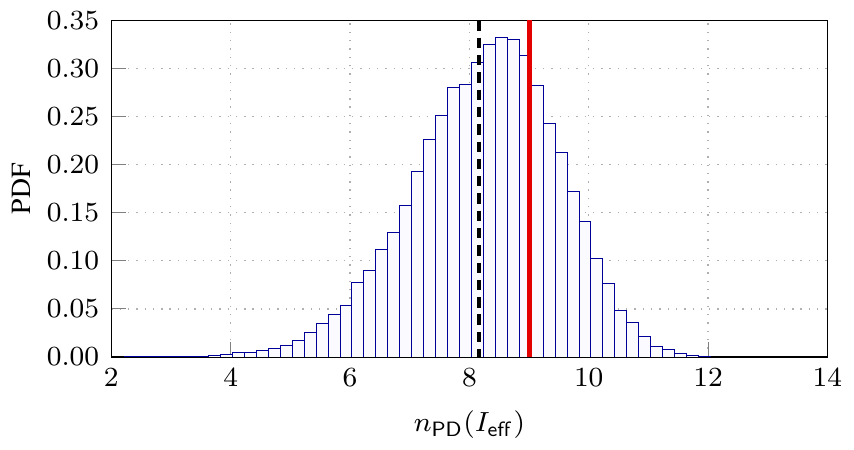}
    \vspace{-18pt}
\caption{The PDF of the Ornstein-Uhlenbeck-simulated $\pditer(\ieff)$ for the $(5,10,L,N=1000)$ SC-LDPC code ensemble under BP decoding for $\iterlim = 200$ and $\e = 0.47$.}
\label{fig:r1_ou}
    \vspace{-2ex}
\end{figure}

In summary, we need to approximate the distribution of $\pditer(\ieff)$ and use it in~\eqref{eq:pdf_x_approx}--\eqref{eq:fer_min} to predict the FER.
We provide several ways to model $\pditer(\ieff)$ below.

\subsection{Simulating $\pditer(\ieff)$ Using an Ornstein-Uhlenbeck Process}
\label{sse:fl_jumping_ou}

We can approximate the distribution of $\pditer(\ieff)$ using the following heuristic: 
We keep the assumption that an iteration of BP corresponds to as many iterations of peeling decoding as there are degree-one CNs available to decode.
However, we no longer assume the normalized number of available degree-one CNs across iterations during the steady state, $\rprocess$, to be constant.
Instead, we assume that each BP iteration is equivalent to advancing along a peeling decoding trajectory---i.e., a realization of $\rprocess$---by as much as there are degree-one CNs in the current position on the peeling decoding trajectory.
This heuristic model allows us to directly simulate the PDF of $\pditer(\ieff)$ from Monte-Carlo realizations of the Ornstein-Uhlenbeck process that correspond to realizations of the steady-state of $\rprocess$.
Indeed, from each Ornstein-Uhlenbeck realization of $\rprocess$ we can generate a realization of $\pditer(K)$ as
\begin{equation}
    \pditer(K) = \pditer(K - 1) + r_1\left(\pditer(K - 1)\right)\,,
    \label{eq:pditer_ou}
\end{equation}
starting with $\pditer(1) = r_1(0)$.
We remark that when $\rprocess$ is set to be constant and equal to $\rmean$ from~\eqref{eq:rprocess_mean}, $\pditer(K)$ boils back down to $\gammasingle (\estar - \e) K$.

Fig.~\ref{fig:r1_ou} shows an example PDF of $\pditer(\ieff)$ simulated via~\eqref{eq:pditer_ou} from Monte-Carlo realizations of the Ornstein-Uhlenbeck process with parameters $(\gammasingle, \nusingle, \thetasingle)$ in~\eqref{eq:rprocess_mean}--\eqref{eq:rprocess_cov} chosen to match the first two moments of $\rprocess$ (blue histogram).
We observe that the average value of $\pditer(\ieff)$ (vertical black dashed line) is smaller than $\gammasingle (\estar - \e) \ieff$ (vertical red solid line).
Further, the left tail of the PDF is ``heavier'' than the right.
This can be explained by the asymmetry in the way temporal correlation in $\rprocess$ influences $\pditer(K)$ in~\eqref{eq:pditer_ou}: when $r_1\left(\pditer(K-1)\right)$ is large, $\pditer(K)$ will end up far away from $\pditer(K-1)$, so the next increment in $\pditer$ will be statistically close to a sample from an independent Gaussian random variable centered at $\gammasingle (\estar - \e)$.
When $r_1\left(\pditer(K-1)\right)$ is small, on the other hand, the next increment in $\pditer$ will also likely be small.

\begin{figure}[!t]
    \centering
    \includegraphics[width=\columnwidth]{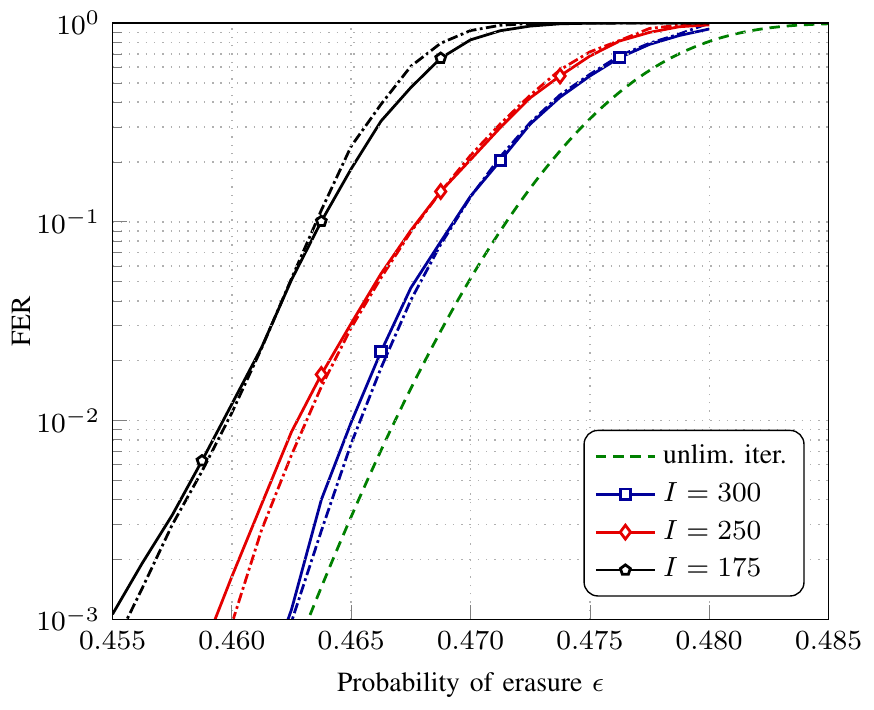}
    \vspace{-18pt}
    \caption{FER for the $(5,10,L=50,N=1000)$ SC-LDPC code ensemble under BP decoding for different limits on the number of iterations $\iterlim$ (solid curves) and its approximation using the PDF of $\pditer(\ieff)$ from~\eqref{eq:pditer_ou} based on the Ornstein-Uhlenbeck process (corresponding dash-dotted curves).}
\label{fig:fl_jumping_ou}
    \vspace{-2ex}
\end{figure}

The first scaling law we propose in this section uses the simulated distribution of $\pditer(\ieff)$ based on the Ornstein-Uhlenbeck model~\eqref{eq:pditer_ou} in~\eqref{eq:pdf_x_approx}--\eqref{eq:fer_min} to estimate the FER.
The corresponding approximations are shown in Fig.~\ref{fig:fl_jumping_ou} for the terminated $(5,10,L=50,N=1000)$ SC-LDPC code ensemble and $\iterlim = \{ 175, 250, 300 \}$ (dash-dotted curves).
The match between the predicted and simulated (solid) FER curves is remarkable---evidently, the iterative model based on the Ornstein-Uhlenbeck process~\eqref{eq:pditer_ou} is a good approximation of the behavior of BP decoding.
The drawback of the model is that it requires Monte-Carlo approximation of the PDF of $\pditer(\ieff)$ for every combination of $(\e, N, \iterlim)$; however, simulating $\pditer(\ieff)$ via~\eqref{eq:pditer_ou} is still far less computationally complex than simulating BP decoding.

\subsection{Gaussian Propagation Distance Model}
\label{sse:fl_q}

In this section, we introduce a model that does not require Monte-Carlo simulation of the distribution of $\pditer(\ieff)$ as in Section~\ref{sse:fl_jumping_ou}.
Instead, the model approximates $\pditer(\ieff)$ by a Gaussian random variable.
To that end, we will rely on the fact that a time integral of an Ornstein-Uhlenbeck process is normally distributed, as we describe below.

\subsubsection{Ornstein-Uhlenbeck process: necessary background}
\label{sss:ou_primer}

A generic Ornstein-Uhlenbeck process $Z_t$ can be described using the following stochastic differential equation:
\begin{equation}
    \der Z_t = - \oudecay (x_t - \oumean) \der t + \sigma \der B_t\,,
    \label{eq:ou_def}
\end{equation}
where $B_t$ is the standard Wiener process, $\oudecay$ and $\sigma$ are real positive constants, and $\oumean$ is real.
The first term on the right-hand side of~\eqref{eq:ou_def} can be conceptualized as a mean-reverting factor and the second term as a random fluctuation. The constants $\oudecay$ and $\sigma$ determine their relative importance.

For a fixed sufficiently large $t$, the distribution of $Z_t$ is Gaussian.
Specifically,
\begin{equation}
    Z_t \sim \mathcal{N} \left( \oumean, \frac{\sigma^2}{2\oudecay} \right)\,.
    \label{eq:ou_dist}
\end{equation}
Moreover, for sufficiently large $t + s$,
\begin{equation}
    \Cov{Z_t, Z_s} = \frac{\sigma^2}{2\oudecay} \exp\left( -\oudecay \left| t - s \right| \right)\,.
    \label{eq:ou_cov}
\end{equation}

Crucially, Ornstein and Uhlenbeck showed that a time-integrated Ornstein-Uhlenbeck process is a Gaussian random variable~\cite{ref:Uhle30}. For $t \grows$,
\begin{equation}
    \int\limits_0^t Z_t \der t \sim \mathcal{N} \left( \oumean t, \frac{\sigma^2}{\oudecay^2}t \right)\,.
    \label{eq:ou_integral}
\end{equation}
This result is the cornerstone of our approximation of $\pditer(\ieff)$ as a Gaussian random variable.

\subsubsection{Ornstein-Uhlenbeck process as a model for BP decoding}
\label{sse:ou_bp_model}

Let $\vprocessbp(t)$ denote the number of new VNs (normalized by $N$) resolved in BP iteration $\ell$, where $t$ is
\begin{equation}
    t \defn \ell \cdot \left( \estar - \e \right)\,.
    \label{eq:norm_time}
\end{equation}
The quantity $\vprocessbp(t)$ can be interpreted as the instantaneous decoding speed, measured in VNs (normalized by $N$) per BP iteration.
Here, the time variable $t$ is normalized by the distance to the BP threshold $\left( \estar - \e \right)$ instead of by $N$ as is the case for $\taupd$, discussed in Section~\ref{sse:old_laws}.
Up to the normalization of time, $\vprocessbp(t)$ is equivalent to $\vprocessbp(\ell)$ introduced in Section~\ref{sec:speed}.
We use different time variables in these two cases to avoid confusion.

The process $\vprocessbp(t)$ was introduced in~\cite{ref:Stin16_ppd} to provide an alternative finite-length scaling law for SC-LDPC code ensembles that provides less accurate predictions of the FER than the law in~\cite{ref:Soko20} (and even than the law in~\cite{ref:Olmo15}) but does not rely on peeling decoding or mean evolution.
Indeed, as opposed to $\rmean$, $\E{ \vprocessbp(t) }$ can be obtained directly from density evolution, as we discuss in Section~\ref{sec:speed} (see~\eqref{eq:vbp}).
The authors show that $\vprocessbp(t)$ exhibits a steady-state phase; they approximate the first two moments of this process during the steady-state phase as
\begin{align}
    \E{ \vprocessbp(t) } &\approx \gammabp \left( \estar - \e \right)\,, \label{eq:vprocessbp_mean} \\
    \Var{ \vprocessbp(t) } &\approx \frac{\nubp}{N}\,, \label{eq:vprocessbp_var} \\
    \Cov{ \vprocessbp(t), \vprocessbp(s) } &\approx \frac{\nubp}{N} \exp\left( -\thetabp \left| t - s \right| \right)\,. \label{eq:vprocessbp_cov}
\end{align}
We apply the same refinement of the law as we did in~\cite{ref:Soko20} and model the two-wave process for the terminated ensemble as a combination of two independent Ornstein-Uhlenbeck processes.
We therefore estimate the triple $(\gammabp,\nubp,\thetabp)$ from the truncated ensemble, as we did in~\cite{ref:Soko20} for peeling decoding to estimate $(\gammasingle,\nusingle,\thetasingle)$.
(We review the peeling decoding-based laws in Section~\ref{sse:old_laws}.)
The steady-state level constant $\gammabp$ is estimated along with $\iterstart$ and $\iterend$ from density evolution: once $\iterstart$ and $\iterend$ are estimated as described in Section~\ref{sec:speed}, we estimate $\gammabp$ from the average of $\E{ \vprocessbp(\ell) }$ for $\ell \in \left[ \iterstart, \iterend \right]$.
The covariance parameters $\nubp$ and $\thetabp$ are estimated from Monte-Carlo simulations of BP decoding. For our running example of the $(5, 10, N, L)$ SC-LDPC code ensemble, $\nubp$ and $\thetabp$ are estimated for $(N = 5000, \e = 0.465)$ to be $\nubp \approx 0.41, \thetabp \approx 2.74$.
The scaling parameters $\gammabp,\nubp$, and $\thetabp$ are illustrated in Fig.~\ref{fig:de_illustration} along with $\iterstart$ and $\iterend$.

We follow~\cite{ref:Stin16_ppd} and model $\vprocessbp(t)$ in the steady state by an Ornstein-Uhlenbeck process.
Equating the moments~\eqref{eq:vprocessbp_mean}--\eqref{eq:vprocessbp_cov} to those in~\eqref{eq:ou_dist}--\eqref{eq:ou_cov} yields
\begin{align}
    \oumean = \gammabp \left( \estar - \e \right)\,,~ ~ ~ ~\oudecay = \thetabp\,,~ ~ ~ ~\sigma^2 = 2 \thetabp \cdot \frac{\nubp}{N}\,.
    \label{eq:vbprocess_ou_params}
\end{align}

\subsubsection{Normal approximation based on a time-integrated Ornstein-Uhlenbeck process}
\label{sse:bp_normal}

The total number of VNs decoded in $\ieff$ iterations, $\pditer(\ieff)$, can be expressed as a time integral of $\vprocessbp(t)$, divided by $\left( \estar - \e \right)$ to convert the units of speed to VNs per \textit{normalized} iteration,
\begin{align}
    \pditer(\ieff) &\approx \frac{1}{\estar - \e} \int\limits_{\tstart}^{\tstart + \teff} \vprocessbp(t) \der t \nonumber \\
    &= \frac{1}{\estar - \e} \int\limits_0^{\teff} \vprocessbp(t + \tstart) \der t \,, \label{eq:pditer_integral}
\end{align}
where
\begin{align}
    \tstart \defn \iterstart \left( \estar - \e \right) \,, \quad \teff \defn \ieff \left( \estar - \e \right)\,. \label{eq:teff}
\end{align}

We can now use the expression for a time-integrated Ornstein-Uhlenbeck process~\eqref{eq:ou_integral} with appropriately chosen parameters~\eqref{eq:vbprocess_ou_params} to approximate the distribution of $\pditer(\ieff)$ from~\eqref{eq:pditer_integral} as
\begin{align}
    \left( \estar\!-\!\e \right) \pditer(\ieff) &\sim \mathcal{N} \left( \oumean \teff, \frac{\sigma^2}{\oudecay^2} \teff \right) \nonumber \\
    &\stackrel{\text{(a)}}{=} \mathcal{N} \left( \gammabp \left( \estar\!-\!\e \right) \teff, \frac{2 \nubp}{N\thetabp} \teff \right) \label{eq:pditer_derivation} \\
    &\stackrel{\text{(b)}}{=} \mathcal{N} \left( \gammabp \left( \estar\!-\!\e \right)^2 \ieff, \frac{2 \nubp}{N\thetabp} \left( \estar\!-\!\e \right) \ieff \right)\,, \nonumber
\end{align}
\begin{align}
    \pditer(\ieff) &\sim \mathcal{N} \left( \gammabp \left( \estar - \e \right) \ieff, \frac{2 \nubp \ieff}{N\thetabp \left( \estar - \e  \right)} \right) \,,
    \label{eq:pditer_normal_bp}
\end{align}
where in (a) we used~\eqref{eq:vbprocess_ou_params} and in (b) we used~\eqref{eq:teff}.
Notably, the approximation~\eqref{eq:pditer_normal_bp} reveals that both mean and variance of $\pditer(\ieff)$ grow linearly with $\ieff$.
We must also remark that the approximation~\eqref{eq:pditer_normal_bp} does not account for the possibility that the underlying Ornstein-Uhlenbeck process becomes negative.
\begin{figure}[!t]
    \centering
    \includegraphics[width=\columnwidth]{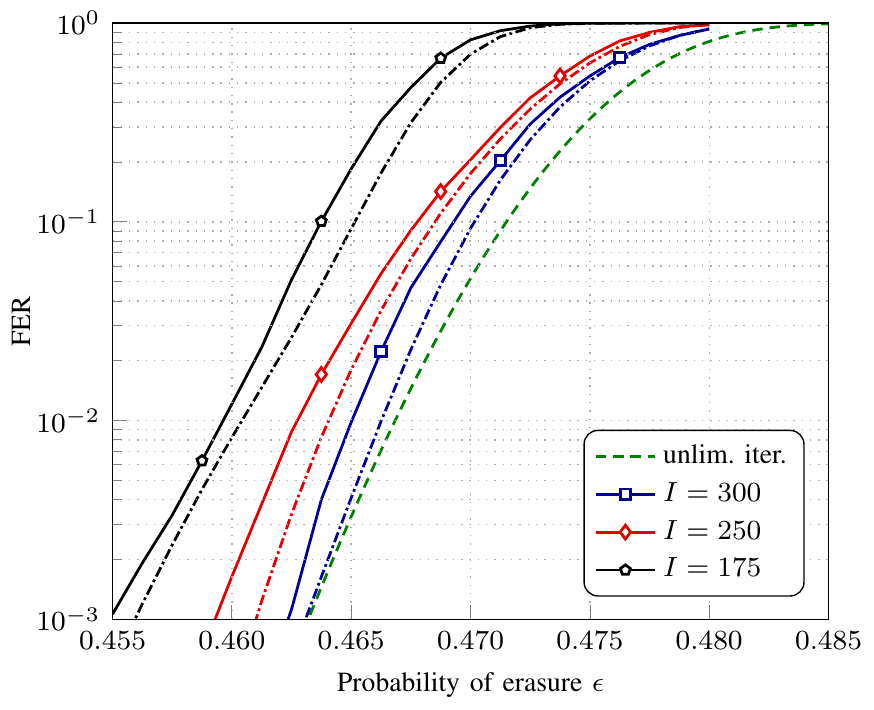}
    \vspace{-18pt}
    \caption{FER for the $(5,10,L=50,N=1000)$ SC-LDPC code ensemble under BP decoding for different limits on the number of iterations $\iterlim$ (solid curves) and its approximation using~\eqref{eq:pdf_x_approx}--\eqref{eq:fer_min} with the distribution of $\pditer(\ieff)$ from~\eqref{eq:pditer_normal_bp} (corresponding dash-dotted curves).}
\label{fig:fl_gaussian_min}
    \vspace{-2ex}
\end{figure}

The second scaling law we propose in this section uses the normal approximation~\eqref{eq:pditer_normal_bp} to the distribution of $\pditer(\ieff)$ in~\eqref{eq:pdf_x_approx}--\eqref{eq:fer_min} to estimate the FER.
The corresponding predictions are shown in Fig.~\ref{fig:fl_gaussian_min} for our running example of the terminated $(5,10,L=50,N=1000)$ SC-LDPC code ensemble and $\iterlim = \{ 175, 250, 300 \}$ (dash-dotted curves).
We observe that, while being a good approximation to the simulated FER, the predictions are more optimistic than the predictions based on the simulated Ornstein-Uhlenbeck process from Section~\ref{sse:fl_jumping_ou} (cf.~Fig.~\ref{fig:fl_jumping_ou}).

\subsubsection{Shifted normal approximation}
\label{sse:shifted_normal}

\begin{figure}[!t]
    \centering
    \includegraphics[width=\columnwidth]{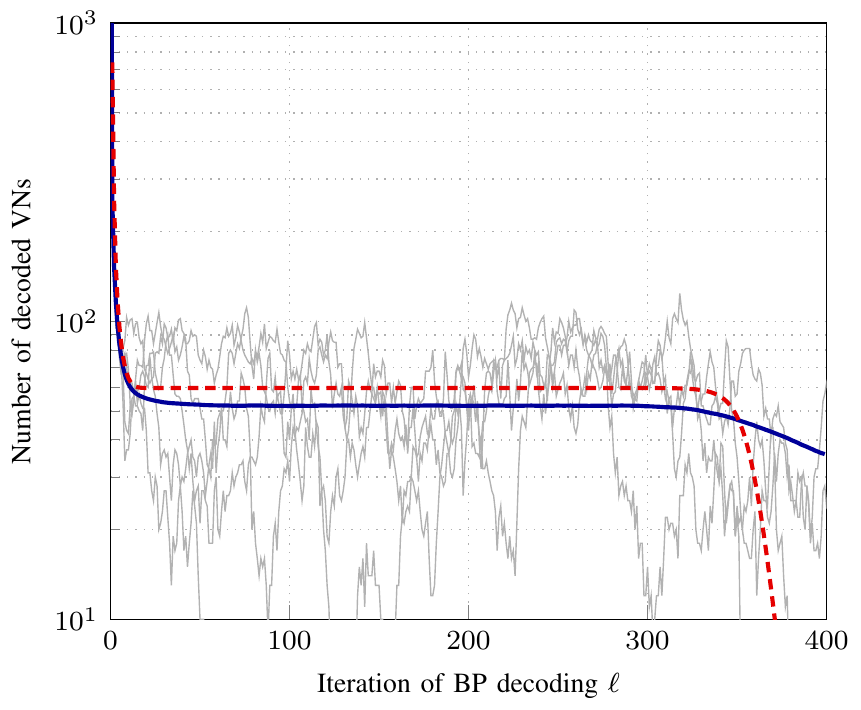}
    \vspace{-18pt}
    \caption{The number of VNs decoded per BP iteration for the truncated $(5,10,L=50,N=1000)$ SC-LDPC code ensemble and $\e = 0.47$: the simulated average from $10^5$ trajectories (blue solid curve) and the mean from density evolution (red dashed curve).
    Several simulated trajectories are shown as thin gray lines.}
\label{fig:traj_gamma_gap}
    \vspace{-2ex}
\end{figure}

Part of the reason behind the mismatch between the simulated and predicted FER in Fig.~\ref{fig:fl_gaussian_min} is that~\eqref{eq:vprocessbp_mean} overestimates the average number of VNs decoded in a BP iteration for finite $N$, as shown in Fig.~\ref{fig:traj_gamma_gap} for the truncated $(5,10,L=50,N=1000)$ SC-LDPC code ensemble and $\e = 0.47$.
\begin{figure}[!t]
    \centering
    \includegraphics[width=\columnwidth]{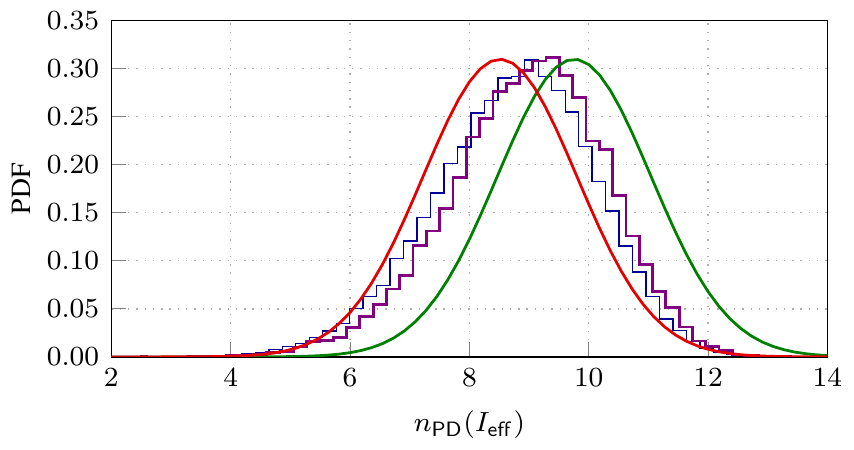}
    \vspace{-18pt}
    \caption{The PDF of the total number of VNs decoded during the steady state for the truncated $(5,10,L=50,N=1000)$ SC-LDPC code ensemble, $\e = 0.47$ and $\iterlim = 200$: the simulated histogram (thin blue curve), the Gaussian approximation~\eqref{eq:pditer_normal_bp} with density evolution- (green curve) or BP simulation-based (red curve) mean, and the simulated PDF based on the iterative Ornstein-Uhlenbeck model from Section~\ref{sse:fl_jumping_ou} (purple curve).}
\label{fig:pditer_hist}
    \vspace{-2ex}
\end{figure}
For the same ensemble and $\epsilon$, Fig.~\ref{fig:pditer_hist} shows that the gap in the steady-state level translates into a shift in the location of the Gaussian approximation~\eqref{eq:pditer_normal_bp} to $\pditer(\ieff)$ (green curve) relative to the simulated histogram (thin blue curve).
Estimating the gap without resorting to Monte-Carlo simulations proves to be difficult; moreover, even if we adjust the location of the Gaussian by shifting its mean to account for the gap, the resulting distribution (red curve) lags behind the simulated histogram (thin blue curve).
Overall, the simulated PDF based on the iterative Ornstein-Uhlenbeck model in Section~\ref{sse:fl_jumping_ou} (purple curve) is the most accurate.
On the other hand, the figure suggests that the two Gaussian models capture the variance of $\pditer(\ieff)$ rather well.

With that in mind, we propose a hybrid model where the Gaussian approximation~\eqref{eq:pditer_normal_bp} is shifted to be located at the average of the iterative Ornstein-Uhlenbeck model~\eqref{eq:pditer_ou}.
Specifically, let $f(\e, N, \ieff)$ denote that average.
Unfortunately, obtaining $f(\e,N,\ieff)$ analytically is challenging; we resort to Monte-Carlo simulations of the iterative Ornstein-Uhlenbeck model~\eqref{eq:pditer_ou} to estimate it.
These simulations indicate that $f(\e, N, \ieff)$ depends linearly on $\ieff$---we can thus estimate the slope of $f(\e, N, \ieff)$ as $\fslope = f(\e, N, \ieff') / \ieff'$ for a specific $\ieff'$ and obtain the values of $f(\e, N, \ieff)$ for other $\ieff$ by scaling that mother curve as $\fslope\ieff$, thereby avoiding the need to re-simulate $f(\e, N, \ieff)$ for every $\ieff$.
This is equivalent to using
\begin{equation}
    \oumean = \fslope
    \label{eq:gamma_jumping_loc}
\end{equation}
instead of $\oumean = \gammabp\left( \estar - \e \right)$ in~\eqref{eq:vbprocess_ou_params}.

Putting it all together, the third and last scaling law we propose in this section approximates the distribution of $\pditer(\ieff)$ as
\begin{align}
    \pditer(\ieff) &\sim \mathcal{N} \left( \fslope\ieff, \frac{2 \nubp \ieff}{N\thetabp \left( \estar - \e  \right)} \right)
    \label{eq:pditer_normal_jumping_loc}
\end{align}
and uses it in~\eqref{eq:pdf_x_approx}--\eqref{eq:fer_min} to estimate the FER.
\begin{figure}[!t]
    \centering
    \includegraphics[width=\columnwidth]{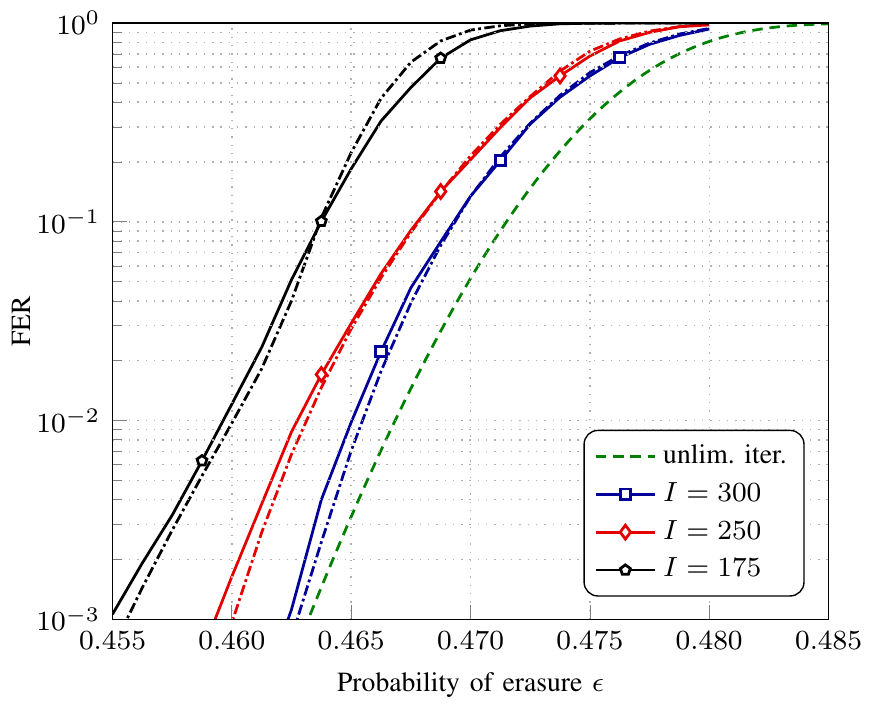}
    \vspace{-18pt}
    \caption{FER for the $(5,10,L=50,N=1000)$ SC-LDPC code ensemble under BP decoding for different limits on the number of iterations $\iterlim$ (solid curves) and its approximation using~\eqref{eq:pdf_x_approx}--\eqref{eq:fer_min} with the distribution of $\pditer(\ieff)$ from~\eqref{eq:pditer_normal_jumping_loc} (corresponding dash-dotted curves).}
\label{fig:fl_gaussian_min_jumping_loc}
    \vspace{-2ex}
\end{figure}

The resulting FER predictions for the $(5,10,L=50,N=1000)$ SC-LDPC code ensemble are shown in Fig.~\ref{fig:fl_gaussian_min_jumping_loc} for $\iterlim = \{ 175, 250, 300 \}$ (dash-dotted curves).
We chose $\ieff'=350 - \iterstart - \iterend$, estimated $\fslope = f(\e, N=1000, \ieff') / \ieff'$ for several $\e$, and linearly interpolated it to obtain the intermediate values.
(We remark that $\iterstart$ and $\iterend$ depend on $\e$ in their turn; as we discuss in Section~\ref{sec:speed}, we also estimate them for several $\e$ and use linear interpolation to obtain the intermediate values.)
The resulting predictions are very close to those in Fig.~\ref{fig:fl_jumping_ou} by the iterative Ornstein-Uhlenbeck model (i.e., by the first scaling law proposed in this section).

We conclude that the hybrid model provides an interesting trade-off between accuracy and analytical tractability.
Importantly, unlike the first scaling law in this section, it does not require Monte-Carlo simulation for every combination of $(\e, N, \iterlim)$---estimating $\fslope$ from a single value of the number of BP iterations allows us to obtain the whole family of FER curves for different limits on the number of BP iterations.

\begin{figure}[!t]
    \centering
    \includegraphics[width=\columnwidth]{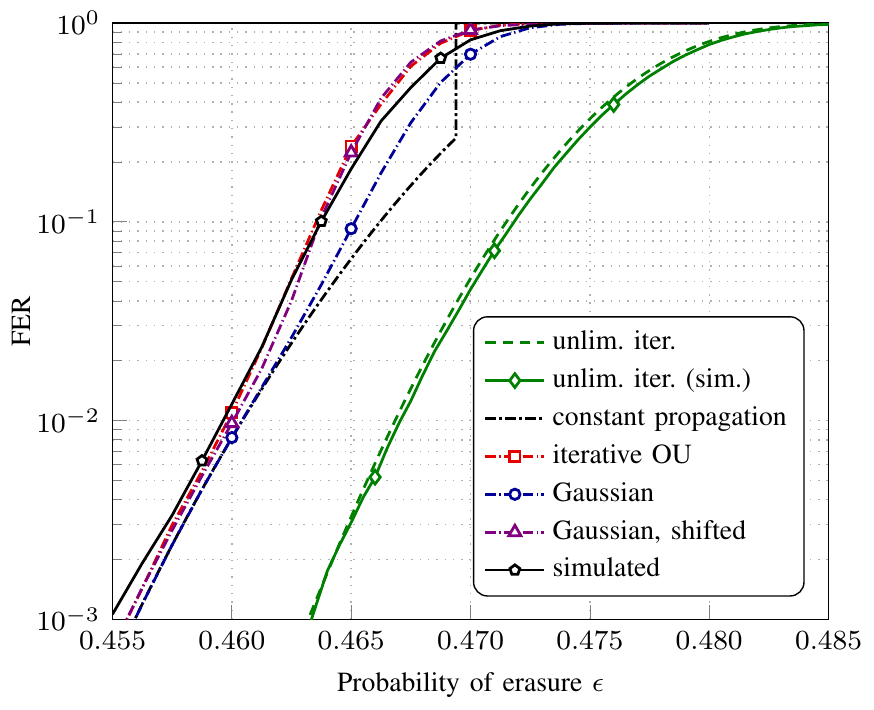}
    \vspace{-18pt}
    \caption{Simulated FER for the $(5,10,L=50,N=1000)$ SC-LDPC code ensemble under BP decoding for $\iterlim = 175$ (solid curve) and its approximations (dash-dotted curves).}
\label{fig:fl_lim_iter_summary_i175}
    \vspace{-2ex}
\end{figure}

\subsection{Discussion}

Before we tackle the FER prediction for sliding window decoding with a limit on the number of iterations in Section~\ref{sec:sw_lim_iter}, let us summarize the performance-complexity trade-offs associated with the four scaling laws we proposed for full BP decoding.
To that end, Fig.~\ref{fig:fl_lim_iter_summary_i175} groups the simulated (black solid curve with pentagons) and predicted FER curves for the terminated $(5,10,L=50,N=1000)$ SC-LDPC code ensemble under full BP decoding with $\iterlim=175$ using the constant propagation model~\eqref{eq:fl_speed} (black dash-dotted curve) and the randomized propagation model~\eqref{eq:pdf_x_approx}--\eqref{eq:fer_min} with the approximated distribution of $\pditer(\ieff)$ based on the iterative Ornstein-Uhlenbeck model~\eqref{eq:pditer_ou} (red dash-dotted curve with squares), on the normal distribution for the integrated Ornstein-Uhlenbeck process~\eqref{eq:pditer_normal_bp} (blue dash-dotted curve with circles), and on the shifted normal distribution~\eqref{eq:pditer_normal_jumping_loc} (purple dash-dotted curve with triangles).
For reference, we have also included the simulated FER curve for unlimited number of iterations (green solid curve with diamonds) alongside our prediction from~\cite{ref:Soko20} (given in~\eqref{eq:our_fer}) (green dashed curve).

Besides the scaling parameters also required for the scaling law with unlimited iterations~\eqref{eq:our_fer}, the black dash-dotted prediction curve obtained via the constant propagation model~\eqref{eq:fl_speed} requires calculating $\tmin$, which is a function of quantities we estimate from mean and density evolution, as we describe in Section~\ref{sec:fl_constant_speed}.
In that sense, the constant propagation model~\eqref{eq:fl_speed} is no more computationally challenging than the scaling law for unlimited iterations~\eqref{eq:our_fer}.
On the other hand, the constant propagation model does not capture well the transition regions of the FER curve, as exemplified by the discontinuous jump of the black dash-dotted curve around $\e = 0.47$ and discussed in Section~\ref{sec:fl_constant_speed}.

The iterative Ornstein-Uhlenbeck model~\eqref{eq:pditer_ou} yields very accurate FER predictions (red dash-dotted curve with squares); however, it requires Monte-Carlo simulation of the Ornstein-Uhlenbeck process and numerical approximation of the PDF of $\pditer(\ieff)$ from the simulated realizations for every $(\e, N, \iterlim)$.
The normal approximation based on the time-integrated Ornstein-Uhlenbeck process~\eqref{eq:pditer_normal_bp}, shown in Fig.~\ref{fig:fl_lim_iter_summary_i175} as the blue dash-dotted curve with circles, is computationally less complex than the iterative Ornstein-Uhlenbeck model once the parameters $(\gammabp, \nubp, \thetabp)$ are estimated, since it does not require estimating a probability distribution via Monte-Carlo simulation.
However, it is also less accurate.
The accuracy can be improved by shifting the Gaussian to be located at the mean of the iterative Ornstein-Uhlenbeck model, which we describe in Section~\ref{sse:shifted_normal} to obtain the purple dash-dotted curve with triangles in Fig.~\ref{fig:fl_lim_iter_summary_i175} via~\eqref{eq:pditer_normal_jumping_loc}.
The shifted normal approximation is less computationally complex than the iterative Ornstein-Uhlenbeck model---not only does it rely on Gaussian distribution and thus simplify numerical integration in~\eqref{eq:fer_min}, but it also avoids simulating the iterative Ornstein-Uhlenbeck model~\eqref{eq:pditer_ou} for every $\iterlim$\textemdash while achieving similar accuracy.

\section{Finite-Length Scaling: Sliding Window Decoding with a Limited Number of Iterations}
\label{sec:sw_lim_iter}

The core idea behind the scaling laws in Section~\ref{sec:fl_pditer} is to model the number of bits recovered by a given number of BP iterations as a time integral of the Ornstein-Uhlenbeck process $\vprocessbp(t)$.
In this section, we further develop this approach and use it to estimate the FER under sliding window decoding.
First, however, we need to consider the specific ways in which a limit on the number of iterations affects the probability of decoding error, which we discuss in the two following subsections.

\subsection{Competition Between the Left Wave and the Sliding Window}
\label{sse:left_wave}

In addition to the potential failure of the decoding waves that we analyzed in~\cite{ref:Soko20}, a limit on the number of BP iterations in sliding window decoding introduces a kind of race between the left decoding wave and the sliding window.
Let $\posprocessbpint(\ell) \in \left[0, L - 1\right]$ denote the position of the leftmost VN that remains unrecovered by iteration $\ell$.
We will refer to $\posprocessbpint(\ell)$ as the position of the left wave at iteration $\ell$.
Analogously, let $W_\mathsf{L}(\ell)$ denote the leftmost position within the sliding window, and $W_\mathsf{R}(\ell) = W_\mathsf{L}(\ell) + W$ the next position just outside the right boundary of the sliding window.
If, at any iteration $\ell$, the left boundary of the window overtakes the wave, decoding fails, even though it could potentially have succeeded had the number of iterations not been limited.
Let $O$ denote this overtaking event,
\begin{equation}
    O \defn \big\{\posprocessbpint(\ell) < W_\mathsf{L}(\ell)~\text{for some}~\ell \in [1, \iterlim]\,\big\}\,.
    \label{eq:overtake}
\end{equation}
A necessary condition for successful decoding is that $O$ does not happen.

The setup we are considering can be illustrated by means of the following analogy.
Suppose a user is watching a video via a streaming service.
The video is downloading at a rate that fluctuates around a certain average.
After an initial buffering period, the device starts playing the video.
Then the event $O$ in question is that the buffer is exhausted (and thus the video frozen) at least once during playback.

\subsection{Reduced Maximum Propagation Distance for the Right Wave}
\label{sse:right_wave}

Apart from the possibility of the window overtaking the left wave, the limit on the number of iterations also affects decoding by reducing the maximum distance that can be possibly traveled by the right wave (once the sliding window reaches the right boundary of the chain).
When the number of decoding iterations is not limited, we assume that the right wave can travel by up to $W$ positions, which is reflected in our scaling law for sliding window decoding~\eqref{eq:PfSWb} proposed in~\cite{ref:Soko20}.
Here, we account for the presence of a limit on the number of iterations by estimating the maximum number of positions the right wave can travel to the left while the sliding window is moving to the right, which we denote by $W' \leq W$.

We assume the following three-phase process: In the first phase, which begins when the sliding window starts to cover the last VN position (i.e., the right boundary of the window has reached the end of the coupled chain) and lasts $\iterstart$ iterations, the right wave forms.
Then, in the second phase, the right wave and the window move toward each other.
Finally, in the third phase, which lasts $\iterend$ iterations, the two decoding waves meet and collapse.
During the first and third phase the right wave does not travel, whereas the window moves to the right with speed $V_\mathsf{W} \defn 1 / I_\mathsf{s}$.
During the second phase, the right wave travels leftward with speed $\speedbp$, and the window moves rightward with speed $V_\mathsf{W}$.
The maximum number of positions the right wave can travel, $W'$, corresponds to the distance covered by the right wave in the second phase.

During the first and third phase, the window slides by
\begin{equation}
    \left(\iterstart + \iterend \right) V_\mathsf{W}
\end{equation}
positions while the right wave does not move.
The remaining
\begin{equation}
    W - (\iterstart + \iterend) V_\mathsf{W}
\end{equation}
positions will be covered by the window and the wave jointly with speed $\speedbp + V_\mathsf{W}$, which will take
\begin{equation}
    \Big[ W - (\iterstart + \iterend) V_\mathsf{W} \Big] \cdot \frac{1}{\speedbp + V_\mathsf{W}}
\end{equation}
iterations.
During that time, the right wave will have traveled by
\begin{equation}
    W' \defn \Big[ W - (\iterstart + \iterend) V_\mathsf{W} \Big] \cdot \frac{\speedbp}{\speedbp + V_\mathsf{W}}
    \label{eq:adjusted_w}
\end{equation}
positions, which is the adjusted size of the sliding window we use in our model below.
Naturally, as the limit on the number of iterations is relaxed, $V_\mathsf{W} \to 0$ in~\eqref{eq:adjusted_w} and therefore $W' \to W$.

\subsection{General Form of the Scaling Law}
\label{sse:scaling_law}

We incorporate the limit on the number of iterations during sliding window decoding into our model via the two effects outlined above, namely, the possibility that the window overtakes the left wave (an event denoted by $O$) and the reduction of the maximum distance the right wave can travel.
In essence, successful decoding requires that (\textit{i}) the overtaking $O$ does not happen; (\textit{ii}) the left wave does not run out of VNs to decode while propagating through the first $L - W'$ positions; and (\textit{iii}) the left and right wave jointly cover the last $W'$ positions.
Conditions (\textit{ii}) and (\textit{iii}) are equivalent to the two-phase model we proposed in~\cite{ref:Soko20} to estimate the FER under unlimited iterations as in~\eqref{eq:PfSWb}; the only adjustment is that the size of the second phase is reduced from $W$ to $W'$, defined in~\eqref{eq:adjusted_w}, to account for the reduction of the maximum reach of the right wave, as discussed in more detail in Section~\ref{sse:right_wave}.
Condition~(\textit{i}) is introduced in Section~\ref{sse:left_wave}.

Modeling the three conditions as independent events, we approximate the FER under sliding window decoding with a limit on the number of BP iterations as
\begin{align}
    \FER &= 1 -
    \Big( 1 - \operatorname{Pr}\left\{O\right\} \Big) \nonumber \\
    &\phantom{=} \cdot \left(1-P_{\mathsf{f,u}}^{(L-W')}\right)
    \left(1-P_{\mathsf{f,t}}^{(W')}\right)\,, \label{eq:fer_sw_lim_iter}
\end{align}
where $P_{\mathsf{f,u}}^{(L-W')}$ and $P_{\mathsf{f,t}}^{(W')}$ are the estimated FERs for the unterminated and terminated SC-LDPC code ensembles of length $L - W'$ and $W'$, and are defined in~\eqref{eq:our_fer} and~\eqref{eq:untF}, respectively.

The rest of the section is devoted to the estimation of $\operatorname{Pr}\left\{O\right\}$, the core of our finite-length scaling law for sliding window decoding with a limited number of iterations.

\subsection{Modeling the Race Between the Left Wave and the Window}
\label{sse:race_model}

The general idea behind our approach is to model the stochastic process associated with the position of the left wave, $\posprocessbpint(\ell)$, by a scaled time integral of the Ornstein-Uhlenbeck process that corresponds to $\vprocessbp(\ell)$, the normalized number of bits recovered in iteration $\ell$ defined in Section~\ref{sec:speed}, with an additional noise term.
The wave cannot overtake the right boundary of the window, $W_\mathsf{R}(\ell)$, because no BP iterations are performed there.
On the other hand, if the wave is itself overtaken by the left boundary, $W_\mathsf{L}(\ell)$, decoding is bound to fail.
This motivates us to incorporate the boundaries of the window into our model as an absorbing barrier at $W_\mathsf{L}(\ell)$ and a reflecting barrier at $W_\mathsf{R}(\ell)$.
When a stochastic process hits an absorbing barrier, it remains absorbed indefinitely.
When a stochastic process hits a reflecting barrier, it is reflected back inside the domain~\cite{ref:Pavl14}.
The overtaking event $O$ that we introduced in Section~\ref{sse:left_wave} in~\eqref{eq:overtake} corresponds to $\posprocessbpint(\ell)$ having been absorbed by iteration $\iterlim$.
The probability of this event can be estimated by tracking the evolution of the PDF of $\posprocessbpint(\ell)$ across iterations.
That evolution can in turn be described by a partial differential equation called the Fokker-Planck equation~\cite{ref:Pavl14}.
We numerically solve the initial value problem for the Fokker-Planck equation for $\posprocessbpint(\ell)$ with the boundary conditions that correspond to an absorbing barrier at $W_\mathsf{L}(\ell)$ and a reflecting barrier at $W_\mathsf{R}(\ell)$ and obtain the estimation of the probability of $O$.

\subsubsection{Position of the left wave as an integrated Ornstein-Uhlenbeck process}
\label{sse:scaled_iou_model}

The scaling law for full BP decoding proposed in Section~\ref{sec:fl_pditer} uses a time integral of the Ornstein-Uhlenbeck process $\vprocessbp(\ell)$ as a model for the total number of VNs (normalized by $N$) decoded in a given number of BP iterations.
In the context of sliding window decoding, we are interested instead in the \textit{position} of the wave after a number of iterations.
The basic element of our model is the conversion of the normalized number of VNs decoded in BP iteration $\ell$, $\vprocessbp(\ell)$, to the number of positions traveled by the wave in that iteration.
We assume that $N$ VNs decoded during the steady state of BP decoding advance the wave by $\speedpd$ positions---i.e., by the same number of positions as for peeling decoding~\eqref{eq:calc_speed}.
The number of positions traveled in iteration $\ell$ is then
\begin{equation}
    \vprocessbp(\ell) \speedpd
\end{equation}
and the \textit{total} number of positions traveled in $\ell$ steady-state BP iterations is
\begin{equation}
    \posprocessbpint(\ell) = \pditer(\ell) \speedpd \,,
    \label{eq:sw_integral_basic}
\end{equation}
where $\pditer(\ell)$ is number of VNs decoded in $\ell$ steady-state BP iterations that we introduced in Section~\ref{sec:fl_pditer}.
We can use the model~\eqref{eq:pditer_integral} of $\pditer(\ell)$ in~\eqref{eq:sw_integral_basic} to obtain
\begin{align}
    \posprocessbpint(\ell) = \pditer(\ell) \speedpd = \frac{\speedpd}{\estar - \e} \int\limits_{0}^{\ell \left( \estar - \e \right)} \vprocessbp(t) \der t\,.
    \label{eq:sw_integral_vprocessbp}
\end{align}
For convenience, we have assumed here that iteration \mbox{$\ell = 0$} corresponds to the beginning of the steady state when the position of the wave is zero.

To declutter notation, we do not use the time $t$ normalized by the distance to the threshold~\eqref{eq:norm_time} as we do in Section~\ref{sec:fl_pditer} and in~\eqref{eq:sw_integral_vprocessbp}.
Instead, we use a continuous version of the variable $\ell$ with a unit of time also measured in BP iterations, denoted by $\tau$.
The model~\eqref{eq:sw_integral_vprocessbp} can be rewritten in terms of $\tau$ as
\begin{equation}
    \posprocessbpint(\ell) = \int\limits_0^{\ell} \vprocessbp(\tau)\speedpd\,\der \tau\,.
    \label{eq:pos_sw_lim_iter_integral}
\end{equation}

As we do for $\vprocessbp(t)$ in Section~\ref{sse:ou_bp_model}, we model $\vprocessbp(\tau)$ by an Ornstein-Uhlenbeck process of the form~\eqref{eq:ou_def}.
The parameters $\oudecay$ and $\sigma$ need to be rescaled relative to those in~\eqref{eq:vbprocess_ou_params}, resulting in
\begin{equation}
\begin{aligned}
    \oumean &= \fslope\,,\quad \oudecay = \thetabp\left( \estar - \e \right)\,,\\
    \sigma^2 &= 2 \thetabp \left( \estar - \e \right) \frac{\nubp}{N}\,; \label{eq:vbprocess_tau_ou_params}
\end{aligned}
\end{equation}
the two models (i.e., the one with rescaled time $t$ and the one with rescaled $\oudecay$ and $\sigma$) are equivalent and yield identical predictions.
We remark that in this section we use the same value for the average steady-state level $\oumean$ of $\vprocessbp(\tau)$ as in the shifted normal approximation~\eqref{eq:gamma_jumping_loc}.

The scaling law in Section~\ref{sec:fl_pditer} requires the probability distribution of the time integral of an Ornstein-Uhlenbeck process for a specific $t$, which is known to be Gaussian~\eqref{eq:ou_integral}.
The analysis of sliding window decoding requires a more granular approach, since we need to track the evolution of $\posprocessbpint(\ell)$ over iterations.
In other words, we need to treat $\posprocessbpint(\ell)$ as a stochastic process in its own right.
This is complicated by the fact that an integrated Ornstein-Uhlenbeck process of the form~\eqref{eq:ou_integral} or~\eqref{eq:pos_sw_lim_iter_integral} is not Markov.
However, the two-dimensional process
\begin{equation}
    \left( \posprocessbp(\tau) = \vprocessbp(\tau)\speedpd\,,~ \posprocessbpint(\tau) = \int\limits_0^{\tau} \posprocessbp(s)\,\der s \right)
    \label{eq:two_dim_def}
\end{equation}
is Markov and can be analyzed using standard tools for diffusion processes~\cite{ref:Pavl14}.
(A diffusion process can informally be thought of as a continuous-time Markov processes with continuous sample paths~\cite{ref:Pavl14}.)
The corresponding stochastic differential equation is~\cite[Eq.~(35)]{ref:Bene13}
\begin{equation}
\left\{
\begin{aligned}
    \der \posprocessbp(\tau) &= - \oudecay (\posprocessbp(\tau) - \oumean) \der \tau + \sigmaou \der \Wou \\
    \der \posprocessbpint(\tau) &= \posprocessbp(\tau) \der \tau\,,
\end{aligned}
\right.
\label{eq:iou_sde}
\end{equation}
where $\Wou$ is the standard Wiener process.
To account for the scaling of the Ornstein-Uhlenbeck process $\vprocessbp(\tau)$ (defined via~\eqref{eq:ou_def} with parameters~\eqref{eq:vbprocess_tau_ou_params}) by $\speedpd$ in~\eqref{eq:two_dim_def}, $\oumean$ and $\sigmaou$ in~\eqref{eq:iou_sde} should be rescaled relative to those in~\eqref{eq:vbprocess_tau_ou_params}, resulting in
\begin{equation}
\begin{aligned}
    \oumean &= \fslope \speedpd\,,\quad \oudecay = \thetabp\left( \estar - \e \right)\,,\\
    \sigmaou^2 &= 2 \thetabp \left( \estar - \e \right) \speedpd^2 \frac{\nubp}{N}\,.
\label{eq:x_tau_ou_params}
\end{aligned}
\end{equation}
The need for a subscript in the diffusion coefficient $\sigmaou$ will become apparent in the next subsection.

\subsubsection{Additional diffusion of the left wave's position}
\label{sse:additional_diffusion}

\begin{figure}[!t]
    \centering
    \includegraphics[width=\columnwidth]{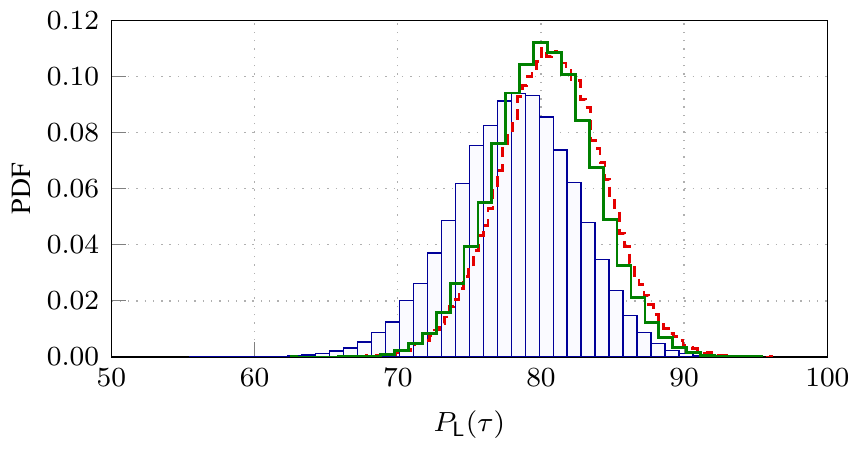}
    \vspace{-18pt}
    \caption{The histogram of the positions of the left wave $\posprocessbpint(\tau)$ after $\tau=412$ BP iterations for the $(5,10,L,N=1000)$ SC-LDPC code ensemble with $\e = 0.455$. Direct Monte-Carlo simulation of the decoding process (blue bars), numerical simulation of the integrated Ornstein-Uhlenbeck process~\eqref{eq:iou_sde}--\eqref{eq:x_tau_ou_params}: continuous (red dashed curve) and discretized (green solid curve) position. The difference in variance between the blue and green histograms is clear.}
\label{fig:erpos_sim}
    \vspace{-2ex}
\end{figure}

To answer whether the integrated Ornstein-Uhlenbeck process \eqref{eq:iou_sde}--\eqref{eq:x_tau_ou_params} is a good model for the propagation distance of the left wave, it can be tested against direct Monte-Carlo simulation of the decoding process.
Fig.~\ref{fig:erpos_sim} compares the distribution of $\posprocessbpint(\ell)$ obtained via direct simulation (blue filled histogram) with the one estimated by simulating $\posprocessbpint(\tau)$ (red dashed curve) and $\lfloor \posprocessbpint(\tau) \rfloor$ (green solid curve) using~\eqref{eq:iou_sde}--\eqref{eq:x_tau_ou_params}.
The integrated Ornstein-Uhlenbeck model captures the average position of the wave relatively well (the means of the blue and green histogram are approximately $78$ and $80$ positions, respectively, after $\ell\!=\!412$ BP iterations) but underestimates its variance.
In other words, the uncertainty in the wave's position does not come solely from the variation in the number of decoded VNs; there must be other factors at play.

We model this additional source of uncertainty via an additional diffusion term that affects $\posprocessbpint(\tau)$ directly.
Throughout iterations, even if the number of decoded VNs is known, the \textit{positions} of these VNs is subject to random fluctuation.
As Fig.~\ref{fig:erpos_sim} shows, by time $\tau$ that fluctuation accumulates into a difference in variance between simulated $\posprocessbpint(\ell)$ and $\lfloor \posprocessbpint(\tau) \rfloor$ from the model~\eqref{eq:iou_sde}--\eqref{eq:x_tau_ou_params}.
We denote the variance of the simulated $\posprocessbpint(\ell)$ by $\sigma^2_\mathsf{sim}(\ell)$ and that of $\lfloor \posprocessbpint(\tau) \rfloor$ from~\eqref{eq:iou_sde}--\eqref{eq:x_tau_ou_params} by $\sigma^2_\mathsf{model}(\ell)$.
We introduce a parameter $\sigmaiou^2$ that corresponds to additional per-iteration variance in position,
\begin{equation}
    \sigmaiou^2 = \frac{\sigma^2_\mathsf{sim}(\ell) - \sigma^2_\mathsf{model}(\ell)}{\ell}\,,
    \label{eq:sigma_iou}
\end{equation}
and incorporate that additional source of variance into our model by changing~\eqref{eq:iou_sde} to
\begin{equation}
\left\{
\begin{aligned}
    \der \posprocessbp(\tau) &= - \oudecay (\posprocessbp(\tau) - \oumean) \der \tau + \sigmaou \der \Wou \\
    \der \posprocessbpint(\tau) &= \posprocessbp(\tau) \der \tau + \sigmaiou \der \Wiou\,,
\end{aligned}
\right.
\label{eq:iou_sde_augmented}
\end{equation}
where $\Wiou$ is another standard Wiener process independent of $\Wou$.
The rationale behind this model is that the variance of the standard Wiener process grows linearly with time~\cite{ref:Pavl14}; adding the term $\sigmaiou \der \Wiou$ in~\eqref{eq:iou_sde_augmented} results in an additional variance of $\sigmaiou^2 \ell$ in $\posprocessbpint(\ell)$ compared with the model in~\eqref{eq:iou_sde}, thereby canceling the mismatch between the simulated and the modeled variance in~\eqref{eq:sigma_iou} but ``spreading'' this variance evenly across iterations.

We treat $\sigmaiou$ as a scaling parameter that depends on $(\dv,\dc)$ only.
We rely on a set of simulated realizations of the decoding process for a certain triple $(\e, N, \ell)$ to estimate $\sigma^2_\mathsf{sim}$.
For our running example of the $(5,10,L,N)$ SC-LDPC code ensemble, we use $(\e = 0.455, N=1000, \ell=412)$ and obtain $\sigmaiou \approx 0.1179$.

As an aside, we remark that the video streaming analogy we introduced in Section~\ref{sse:left_wave} can be stretched further to include the additional uncertainty we are modeling here.
Indeed, the amount of downloaded data is not the only factor that determines how much time of playback is gained.
This is also affected by, e.g., how much movement occurs in consecutive frames of the video, assuming the video is compressed before transmission.

\subsubsection{Modeling the boundaries of the sliding window}
\label{sse:modeling_boudaries}

We are now ready to incorporate the sliding window into our model.
In essence, we do so by limiting the process $\posprocessbpint(\ell)$ to the range from $W_\mathsf{L}(\ell)$ to $W_\mathsf{R}(\ell)$, i.e., to the range of positions covered by the sliding window at iteration $\ell$.
Indeed, $\posprocessbpint(\ell)$ cannot be larger than $W_\mathsf{R}(\ell)$ because sliding window decoding does not perform any BP iterations there.
Similarly, if $\posprocessbpint(\ell)$ ever becomes smaller than $W_\mathsf{L}(\ell)$, the overtaking $O$ happens and decoding fails.
We account for these effects in our model by introducing an absorbing barrier at $W_\mathsf{L}(\ell)$ and a reflecting barrier at $W_\mathsf{R}(\ell)$.

\begin{figure}[!t]
    \centering
    \includegraphics[width=\columnwidth]{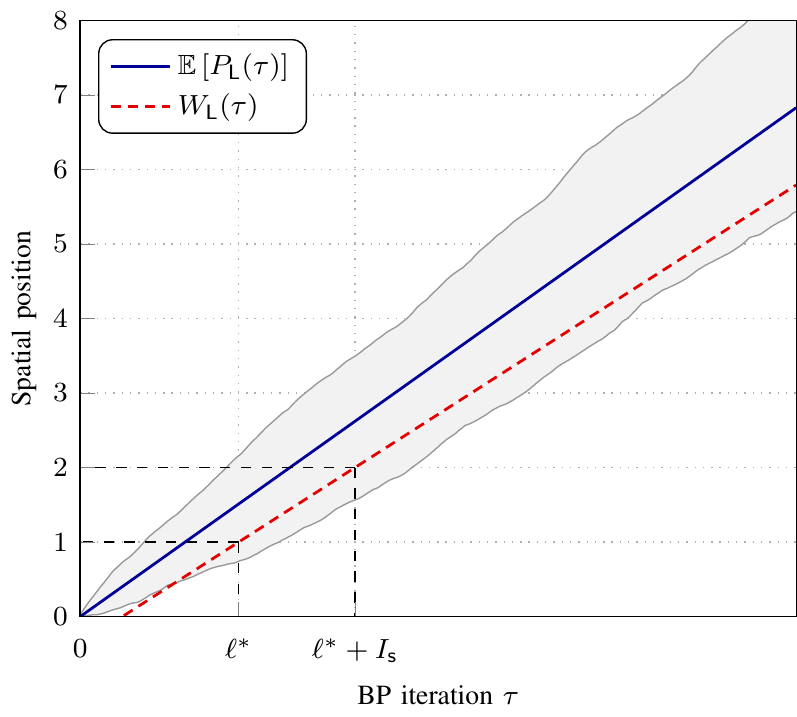}
    \vspace{-5pt}
    \caption{Schematic illustration of the proposed model for the race between the left wave $\posprocessbpint(\tau)$ and the left boundary of the window $W_\mathsf{L}(\tau)$ from~\eqref{eq:smoothed_left_boundary}. The average position of the wave is shown as the blue solid line. The position of the window is shown as the red dashed line. The gray curves represent the minimum and maximum values observed for the position of the wave for a number of simulated realizations of $\posprocessbpint(\tau)$. The range of values between them is shadowed. It is apparent that the spread of $\posprocessbpint(\tau)$ grows with $\tau$. Even though the wave (blue solid line) moves faster than the window (red dashed line) on average, some realizations of $\posprocessbpint(\tau)$ do get overtaken by the window.}
\label{fig:barrier_illustration}
    \vspace{-2ex}
\end{figure}

During decoding, the sliding window moves in discrete steps.
It is convenient to smoothen this movement and assume it to be continuous and linear instead.
We denote the continuously moving boundaries of the window by $W_\mathsf{L}(\tau)$ and $W_\mathsf{R}(\tau)$.
The setup we use is illustrated in Fig.~\ref{fig:barrier_illustration}.
The red dashed line corresponds to $W_\mathsf{L}(\tau)$, which will be included in the model as an absorbing barrier. Its slope is known and equal to $V_\mathsf{W}$.
What remains to be specified is the iteration where $W_\mathsf{L}(\tau)$ crosses $1$, which we denote by $\ell^*$ (see Fig.~\ref{fig:barrier_illustration}).

The value of $\ell^*$ is what couples the model for $\posprocessbpint(\tau)$ with $W_\mathsf{L}(\tau)$.
First, $\ell^*$ depends not only on $I_\mathsf{in}$ but also on the time it takes for the wave to form, $\iterstart$.
A natural choice for $\ell^*$ in that regard would be $I_\mathsf{in} - \iterstart$: it takes $\iterstart$ iterations for the steady state to establish, which eats away from the budget of $I_\mathsf{in}$ initial iterations, and then the window slides to position $1$.
However, numerical simulations show that $\posprocessbpint(\iterstart)$ is already around one, not zero.
In our model, we would like instead to set the initial position of the wave to zero.
We therefore subtract from $\iterstart$ a term that corresponds to $\oumean^{-1}$ in~\eqref{eq:x_tau_ou_params}, which is the average time it takes for the wave to propagate by one position, and set
\begin{equation}
    \ell^* = I_\mathsf{in} - \iterstart + \fslope^{-1}\speedpd^{-1}\,.
\end{equation}

Using our knowledge that $W_\mathsf{L}(\ell^*) = 1$ and that the slope of $W_\mathsf{L}(\tau)$ is $V_\mathsf{W}$, we obtain the continuous version of $W_\mathsf{L}(\ell)$ as
\begin{equation}
    W_\mathsf{L}(\tau) = \tau V_\mathsf{W} + 1 - \ell^* V_\mathsf{W}\,.
    \label{eq:smoothed_left_boundary}
\end{equation}
For simplicity, we ignore the fact that $W_\mathsf{R}(\ell)$ remains flat for the first $\ell^*$ iterations and let
\begin{equation}
    W_\mathsf{R}(\tau) = W_\mathsf{L}(\tau) + W\,.
    \label{eq:smoothed_right_boundary}
\end{equation}

With~\eqref{eq:smoothed_left_boundary}--\eqref{eq:smoothed_right_boundary} we have ensured that the boundaries of the window, $W_\mathsf{L}(\tau)$ and $W_\mathsf{R}(\tau)$, grow linearly with $\tau$.
They correspond to time-dependent absorbing and reflecting barriers and make the domain of $\posprocessbpint(\tau)$ change over time.
We remove that dependency by subtracting the time-dependent term $\tau V_\mathsf{W}$ from $\posprocessbpint(\tau)$.
That corresponds to subtracting $V_\mathsf{W}$ from $\posprocessbp(\tau)$, which can in turn be expressed as a change in $\oumean$.

Putting it all together, we use the model~\eqref{eq:iou_sde_augmented} with
\begin{equation}
\begin{aligned}
    \oumean &= \fslope \speedpd - V_\mathsf{W}\,, \quad \oudecay = \thetabp\left( \estar - \e \right)\,,\\
    \sigmaou^2 &= 2 \thetabp \left( \estar - \e \right) \speedpd^2 \frac{\nubp}{N}\,,
\label{eq:iou_augmented_params}
\end{aligned}
\end{equation}
and the boundaries
\begin{alignat}{2}
    W_\mathsf{L} &= 1 - \ell^* V_\mathsf{W}\,,\quad &\text{absorbing}\,, \label{eq:iou_stable_left_boundary} \\
    W_\mathsf{R} &= W_\mathsf{L} + W\,,\quad &\text{reflecting}\,, \label{eq:iou_stable_right_boundary}
\end{alignat}
which no longer depend on $\tau$.
The initial conditions are $\posprocessbp(0) \sim \mathcal{N} \left( \oumean, \sigmaou^2/(2\oudecay) \right)$ as in~\eqref{eq:ou_dist} and $\posprocessbpint(0) = 0$.

The probability of the overtaking event $O$ corresponds to the probability of $\posprocessbpint(\tau)$ being absorbed at the barrier $W_\mathsf{L}$ by the time $\ell^* + (L - 1) I_\mathsf{s}$.
The latter probability can be obtained by solving the Fokker-Planck equation for $(\posprocessbp(\tau), \posprocessbpint(\tau))$ with appropriately chosen initial and boundary conditions.
The following subsection provides a summary of the associated results for general multidimensional diffusion processes.

\subsubsection{Necessary background on the Fokker-Planck equation}
\label{sse:diffusion_background}

Let $\boldsymbol{Y}_{\tau}$ be a time-homogeneous diffusion process on $\mathbb{R}^{d}$ defined as the solution to the It\^o stochastic differential equation~\cite[Ch. 3]{ref:Pavl14}
\begin{equation}
    \der \boldsymbol{Y}(\tau) = \boldsymbol{b}(\boldsymbol{Y}(\tau)) \der \tau + \boldsymbol{\sigma}(\boldsymbol{Y}(\tau))\der \boldsymbol{B}_\tau\,,
    \label{eq:itosde}
\end{equation}
where $\boldsymbol{B}_\tau$ is the standard Wiener process on $\mathbb{R}^{n}$,
$\boldsymbol{b}(\boldsymbol{Y}) \in \mathbb{R}^{d}$ is a drift vector,
and $\boldsymbol{\sigma}(\boldsymbol{Y}) \in \mathbb{R}^{d \times n}$. The matrix $\boldsymbol{\Sigma}(\boldsymbol{Y}) = \boldsymbol{\sigma}(\boldsymbol{Y}) \boldsymbol{\sigma}(\boldsymbol{Y})^{T} \in \mathbb{R}^{d \times d}$ is known as the diffusion matrix of $\boldsymbol{Y}(\tau)$.
Let the initial condition $\boldsymbol{Y}(0)$ be a random variable with probability density $p_{0}(\boldsymbol{Y})$ independent of $\boldsymbol{B}_\tau$.
Then the probability density $p(\boldsymbol{Y},\tau)$ of $\boldsymbol{Y}(\tau)$ is the solution to the initial value problem for the Fokker-Planck (also known as forward Kolmogorov) equation
\begin{align}
    \frac{\partial p}{\partial \tau} &=\nabla \cdot\left(-\boldsymbol{b}(\boldsymbol{Y}) p+\frac{1}{2} \nabla \cdot(\boldsymbol{\Sigma}(\boldsymbol{Y}) p)\right) \label{eq:fokker_planck_general} \\
    &=-\sum_{i=1}^{d} \frac{\partial}{\partial x_{i}}\left(b_{i}(\boldsymbol{Y}) p\right)+\frac{1}{2} \sum_{i, j=1}^{d} \frac{\partial^{2}}{\partial x_{i} \partial x_{j}}\left(\Sigma_{i j}(\boldsymbol{Y}) p\right)\,, \nonumber \\
    p(\boldsymbol{Y}, 0) &= p_{0}(\boldsymbol{Y})\,, \nonumber
\end{align}
where $x_{\{i,j\}}\,, b_i(\boldsymbol{Y})\,,$ and $\Sigma_{i j}(\boldsymbol{Y})$ denote the components of $\boldsymbol{Y}, \boldsymbol{b}(\boldsymbol{Y})\,,$ and $\boldsymbol{\Sigma}(\boldsymbol{Y}),$ respectively~\cite[Proposition~3.3 and Eq. (4.1)]{ref:Pavl14}.
The Fokker-Planck equation describes the evolution of the probability density of $\boldsymbol{Y}(\tau)$ over time.

Absorbing or reflecting barriers affect $p(\boldsymbol{Y},\tau)$ through additional boundary conditions for the Fokker-Planck equation~\eqref{eq:fokker_planck_general}.
An absorbing barrier $\mathfrak{B}_\mathsf{a}$ imposes a Dirichlet boundary condition on~\eqref{eq:fokker_planck_general}, namely
\begin{equation}
    p(\boldsymbol{Y},\tau) = 0 \quad \forall \boldsymbol{Y} \in \mathfrak{B}_\mathsf{a}\,.
    \label{eq:absorbing}
\end{equation}
Likewise, a reflecting barrier $\mathfrak{B}_\mathsf{r}$ translates into a Neumann boundary condition for~\eqref{eq:fokker_planck_general}. Specifically,
\begin{equation}
    \boldsymbol{n} \cdot \left(\boldsymbol{b}(\boldsymbol{Y}) p - \frac{1}{2} \nabla \cdot(\boldsymbol{\Sigma}(\boldsymbol{Y}) p)\right) = 0 \quad \forall \boldsymbol{Y} \in \mathfrak{B}_\mathsf{r}\,,
    \label{eq:reflecting}
\end{equation}
where $\boldsymbol{n}$ denotes a vector normal to $\mathfrak{B}_\mathsf{r}$.
In other words, the probability density must vanish at the absorbing barrier $\mathfrak{B}_\mathsf{a}$, and there should be no probability flow at the reflecting barrier $\mathfrak{B}_\mathsf{r}$~\cite[pp.~90--91]{ref:Pavl14}.
The probability mass lost by $p(\boldsymbol{Y},\tau)$ is equal to the probability of $\boldsymbol{Y}(\tau)$ having been absorbed at $\mathfrak{B}_\mathsf{a}$ by the time $\tau$,
\begin{equation}
    \operatorname{Pr}\left\{ \boldsymbol{Y}(\tau) \text{~absorbed at~} \mathfrak{B}_\mathsf{a} \right\} = 1 - \int_\Omega p(\boldsymbol{Y},\tau) \der \boldsymbol{Y}\,,
    \label{eq:pabsorbed}
\end{equation}
assuming $\mathfrak{B}_\mathsf{a}$ to be the only absorbing barrier present~\cite[p.~239]{ref:Pavl14}.
The integration in performed over $\Omega$ that denotes the domain of $\boldsymbol{Y}(\tau)$.

\subsubsection{The Fokker-Planck equation for the proposed model}
\label{sse:fokker_planck_wave}

The model~\eqref{eq:iou_sde_augmented} is of the form~\eqref{eq:itosde} with
\begin{equation}
\begin{aligned}
    \boldsymbol{Y}(\tau) &= \left[\begin{array}{c}
\posprocessbp(\tau) \\
\posprocessbpint(\tau) 
\end{array}\right]\,,
\quad
\boldsymbol{b}(\boldsymbol{Y})=\left[\begin{array}{c}
-\oudecay(\posprocessbp - \oumean) \\
\posprocessbp
\end{array}\right]\,, \\
\boldsymbol{\sigma}(\boldsymbol{Y}) &= \left[\begin{array}{ll}
\sigmaou & 0 \\
0 & \sigmaiou
\end{array}\right]\,,
\quad \text { and }
\boldsymbol{B}_\tau=\left[\begin{array}{c}
\Wou \\
\Wiou
\end{array}\right]\,.
\label{eq:sde_notation}
\end{aligned}
\end{equation}

We can therefore use~\eqref{eq:fokker_planck_general}--\eqref{eq:reflecting} to derive the Fokker-Planck equation for the evolution of the PDF $p(\posprocessbp, \posprocessbpint, \tau)$ of the process~\eqref{eq:iou_sde_augmented} as
\begin{align}
    &\frac{\partial p}{\partial \tau} =
    \frac{\partial \oudecay\left( \posprocessbp - \oumean \right)p}{\partial \posprocessbp}
    - \posprocessbp \frac{\partial p}{\partial \posprocessbpint}
    + \frac{\sigmaou^2}{2}\frac{\partial^2 p}{\partial \posprocessbp^2}
    + \frac{\sigmaiou^2}{2}\frac{\partial^2 p}{\partial \posprocessbpint^2}\,,\nonumber\\
    &p(\posprocessbp, \posprocessbpint, 0) = p_{0}(\posprocessbp, \posprocessbpint)\,, \label{eq:fokker_planck}
\end{align}
with boundary conditions
\begin{equation}
\begin{aligned}
    &p\left(\posprocessbp, W_\mathsf{L},\tau\right) = 0\,,\\
    &\posprocessbp p\left( \posprocessbp, W_\mathsf{R}, \tau \right) - \frac{\sigmaiou^2}{2} \left.\frac{\partial p\left( \posprocessbp, \posprocessbpint, \tau\right)}{\partial \posprocessbpint}\right|_{\posprocessbpint = W_\mathsf{R}} = 0\,,
    \label{eq:fokker_planck_boundary_conditions}
\end{aligned}
\end{equation}
where $W_\mathsf{L}$ and $W_\mathsf{R}$ are defined in~\eqref{eq:iou_stable_left_boundary} and \eqref{eq:iou_stable_right_boundary}, respectively, and the parameters $(\oumean,\oudecay,\sigmaou,\sigmaiou)$ are given in~\eqref{eq:iou_augmented_params} and~\eqref{eq:sigma_iou}.

As we specify in Section~\ref{sse:modeling_boudaries}, the initial distribution of $\posprocessbp(\tau)$ and $\posprocessbpint(\tau)$ should be $\posprocessbp(0) \sim \mathcal{N} \left( \oumean, \sigma^2_\mathsf{st} \right)$ and $\posprocessbpint(0) = 0$, where $\sigma^2_\mathsf{st} \defn \sigmaou^2/(2\oudecay)$. 
We should therefore set $p_{0}(\posprocessbp,\posprocessbpint)$ to a PDF whose marginal for $\posprocessbp$ is the PDF of $\mathcal{N} \left( \oumean, \sigma^2_\mathsf{st} \right)$ and for $\posprocessbpint$ the Dirac delta function.
Handling such a distribution numerically is challenging; instead, we set $p_{0}(\posprocessbp,\posprocessbpint)$ to the PDF of the correlated two-dimensional Gaussian distribution
\begin{equation}
    \mathcal{N} \left(
        \left[ \begin{array}{c} \oumean \\ 0 \end{array} \right],
        \left[ \begin{array}{cc}
            \sigma_\mathsf{st}^2 & \rho \delta \sigma_\mathsf{st} \\
            \rho \delta \sigma_\mathsf{st} & \delta^2
        \end{array} \right]
        \right)\,,
    \label{eq:init_condition_normal}
\end{equation}
where $\rho \to 1$ and $\delta \to 0$.
As $\delta \to 0$, the PDF of $\posprocessbpint$ tends to the Dirac delta function as required; the correlation parameter $\rho$ should tend to $1$ because as $\tau \to 0$\,, $\posprocessbp(\tau)$ and its integral $\posprocessbpint(\tau)$ become ever more dependent.
The use of the PDF of~\eqref{eq:init_condition_normal} allows us to avoid the aforementioned numerical issues by backing $\rho$ and $\delta$ off from their limits.

\subsubsection{Numerical solution to the Fokker-Planck equation}
\label{sse:fokker_planck_wave_solution}

To the best of our knowledge, the closed-form solution to the Fokker-Planck equation~\eqref{eq:fokker_planck} in the presence of the boundary conditions~\eqref{eq:fokker_planck_boundary_conditions} is not available.
We therefore resort to solving~\eqref{eq:fokker_planck}--\eqref{eq:fokker_planck_boundary_conditions} numerically using FiPy, a finite-volume solver of partial differential equations~\cite{ref:FiPy09}.
The range of $\posprocessbp$ is limited to $\oumean \pm 4\sigma_\mathsf{st}$ to cover most of the probability mass without overstretching the domain, and that of $\posprocessbpint$ to $[W_\mathsf{L}, W_\mathsf{R}]$.
The resulting rectangular domain is discretized into a regular grid with $\posprocessbp$ and $\posprocessbpint$ split into $200$ and $20W$ segments, respectively.
We set $\rho = 0.99$ and $\delta = 0.1$.
The solver treats~\eqref{eq:fokker_planck} as a convection-diffusion equation; we use implicit convection and diffusion terms~\cite{ref:FiPy09}, which allows us to choose a large time step $1$ without encountering numerical stability issues.

The numerical solution is propagated forward in time until the window is slid through the entire chain, i.e., until
\begin{equation}
    \tau^* \defn \ell^* + (L - 1)I_\mathsf{s}\,.
    \label{eq:total_fp_time}
\end{equation}
The PDF $p(\posprocessbp, \posprocessbpint, \tau^*)$ is used to estimate the probability of the overtaking event $O$ according to~\eqref{eq:pabsorbed} as
\begin{equation}
    \operatorname{Pr}\left\{ O \right\} = 1 -
    \int\limits_{W_\mathsf{L}}^{W_\mathsf{R}}
    \int\limits_{\oumean-4\sigma_\mathsf{st}}^{\oumean+4\sigma_\mathsf{st}}
    p(\posprocessbp, \posprocessbpint, \tau^*) \,\der \posprocessbp\,\der \posprocessbpint\,.
    \label{eq:pr_overtake_fp}
\end{equation}
The estimated $\operatorname{Pr}\left\{ O \right\}$ is then used in~\eqref{eq:fer_sw_lim_iter} to estimate the FER.

We note that our implementation of the solver takes longer to estimate the FER than direct Euler-Maruyama simulation of $10^4$ realizations of~\eqref{eq:iou_sde_augmented} to the same end.
However, the computational complexity of the estimation based on the Fokker-Planck equation does not depend on the FER to attain a given accuracy, which is not the case for the simulation-based approach.
Moreover, no attempt has been made to optimize either implementation.

\subsection{Numerical Results}
\label{sse:sw_lim_iter_results}

\begin{figure}[!t]
    \centering
    \includegraphics[width=\columnwidth]{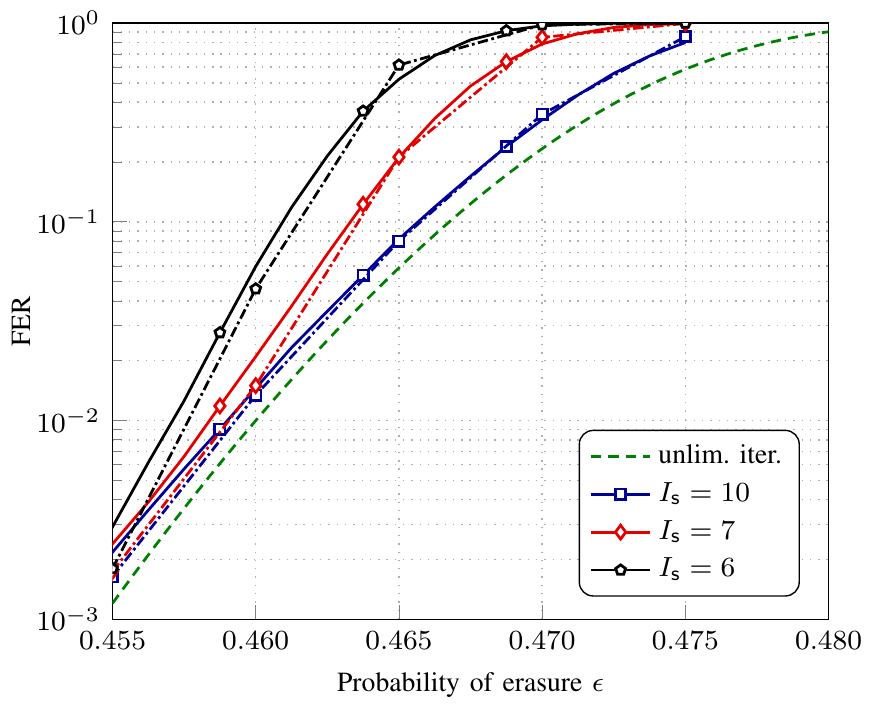}
    \vspace{-18pt}
    \caption{FER for the $(5,10,L=50,N=1000)$ SC-LDPC code ensemble under sliding window decoding with $W=20$ and $I_\mathsf{in}=60$ for different $I_\mathsf{s}$ (solid curves) and its approximation~\eqref{eq:fer_sw_lim_iter} with Fokker-Planck-based estimation of the overtaking probability $O$ (corresponding dash-dotted curves). The green dashed line corresponds to our approximation for the FER with unlimited number of iterations from~\cite{ref:Soko20}.}
\label{fig:fl_sw_lim_iter_in60_1k}
    \vspace{-2ex}
\end{figure}

\begin{figure}[!t]
    \centering
    \includegraphics[width=\columnwidth]{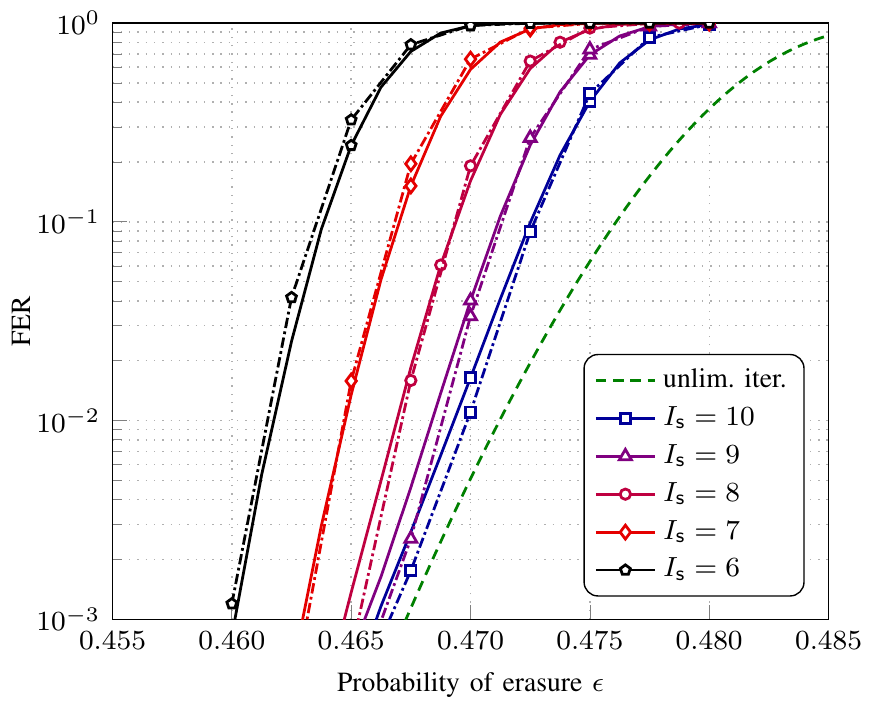}
    \vspace{-18pt}
    \caption{FER for the $(5,10,L=50,N=2000)$ SC-LDPC code ensemble under sliding window decoding with $W=20$ and $I_\mathsf{in}=60$ for different $I_\mathsf{s}$ (solid curves) and its approximation~\eqref{eq:fer_sw_lim_iter} with Fokker-Planck-based estimation of the overtaking probability $O$ (corresponding dash-dotted curves). The green dashed line corresponds to our approximation for the FER with unlimited number of iterations from~\cite{ref:Soko20}.}
\label{fig:fl_sw_lim_iter_in60_2k}
    \vspace{-2ex}
\end{figure}

Fig.~\ref{fig:fl_sw_lim_iter_in60_1k} compares the FER for the $(5,10,L=50,N=1000)$ SC-LDPC code ensemble under sliding window decoding with $W=20$ and $I_\mathsf{in}=60$ for $I_\mathsf{s} = \{ 6,7,10 \}$ (solid curves) with the corresponding approximation using~\eqref{eq:fer_sw_lim_iter}, where the overtaking probability $O$ is estimated using~\eqref{eq:pr_overtake_fp} (corresponding dash-dotted curves).
We observe an impressive match between the simulated and predicted error rates.
A similarly accurate prediction is obtained for $N=2000$ in Fig.~\ref{fig:fl_sw_lim_iter_in60_2k} and for other values of $(\dv, \dc)$ with $\dv > 3$ (not shown).

\begin{figure}[!t]
    \centering
    \includegraphics[width=\columnwidth]{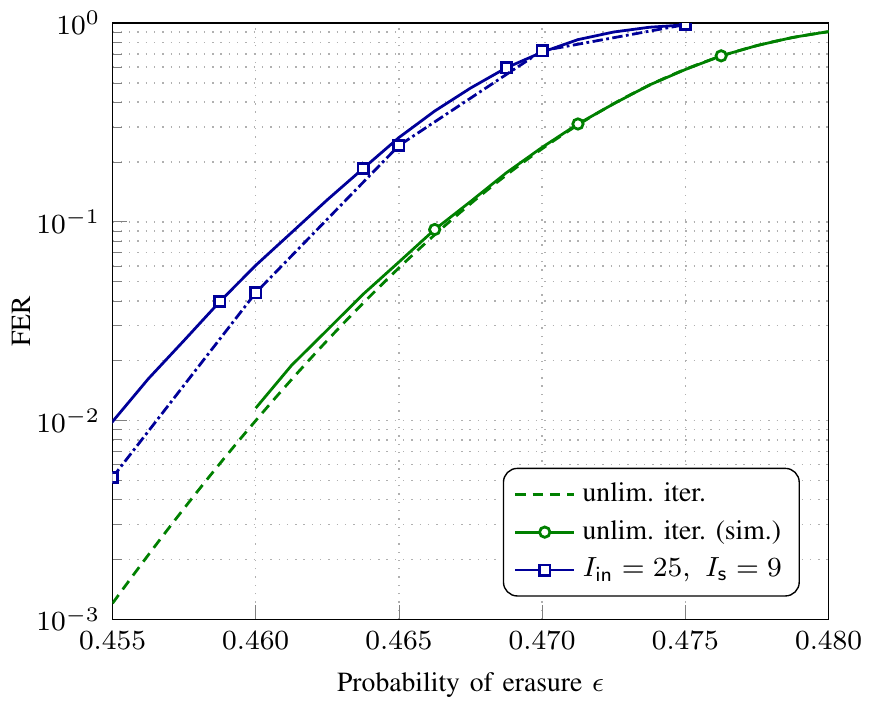}
    \vspace{-18pt}
    \caption{FER for the $(5,10,L=50,N=1000)$ SC-LDPC code ensemble under sliding window decoding with $W=20$, $I_\mathsf{in}=25$, and $I_\mathsf{s} = 9$ (solid curve with squares) and its approximation~\eqref{eq:fer_sw_lim_iter} with Fokker-Planck-based estimation of the overtaking probability $O$ (dash-dotted curve with squares). The green dashed line corresponds to our approximation for the FER with unlimited number of iterations from~\cite{ref:Soko20}.}
\label{fig:fl_sw_lim_iter_in25_i9}
    \vspace{-2ex}
\end{figure}

The quality of the prediction deteriorates for smaller $I_\mathsf{in}$, as exemplified in Fig.~\ref{fig:fl_sw_lim_iter_in25_i9} for $N=1000,I_\mathsf{s}=9$ and $I_\mathsf{in} = 25$.
We observe that degradation when $I_\mathsf{in}$ is such that the left wave is often overtaken at the very beginning of the chain.
This setup, however, is of limited practical relevance---it is sensible for the system designer to ensure that the wave is firmly established by choosing $I_\mathsf{in}$ sufficiently large.
This also lowers the probability that the wave is overtaken by the window early on during decoding if wave propagation happens to slow down for some time.
We have included in Fig.~\ref{fig:fl_sw_lim_iter_in25_i9} the simulated FER curve for unlimited number of iterations (green solid curve with diamonds) alongside our prediction from~\cite{ref:Soko20} using~\eqref{eq:PfSWb} (green dashed curve) for reference.

We remark that we use $\thetabp \approx 2.34$ estimated at $(\e=0.455, N=1000)$ to obtain the predictions in Figs.~\ref{fig:fl_sw_lim_iter_in60_1k}--\ref{fig:fl_sw_lim_iter_in25_i9}.
This is smaller than $\thetabp \approx 2.74$ used in Section~\ref{sec:fl_pditer} and estimated at $(\e = 0.465, N=5000)$.
Choosing the latter value makes the prediction curves slightly more optimistic.
The covariance decay parameter seems to exhibit a stronger dependency on $N$ than in the case of peeling decoding; we leave the investigation of this dependency as a subject for future work.
As a general rule of thumb, one should estimate $\thetabp$ for $(\e, N)$ that lie within the range they are interested in.

\section{Conclusion and Discussion}
\label{sec:conclusion}

The proposed scaling laws for full BP decoding with a limited number of iterations and a scaling law for sliding window decoding with a limited number of iterations provide accurate predictions of the FER.
Modeling the number of bits decoded in a given number of iterations by a time integral of an Ornstein-Uhlenbeck process---the cornerstone of our scaling laws---proves to be a powerful tool in the analysis of decoding schemes with practically relevant constraints on the maximum number of iterations.
More broadly, low-dimensional diffusion processes seem to be able to capture much of the behavior of iterative decoders relevant for error rate prediction.

Another takeaway is that it is important for sliding window decoding to perform a sufficient number of iterations at the beginning of the chain.
It is necessary not only because the decoder must ensure that the decoding wave is established, but also because allowing the decoding wave to propagate further inside the window builds up the decoder's resilience to variation in the wave's propagation speed.

We also remark that the scaling laws we propose in this paper can be extended to predict the bit and block error rate using the same techniques we employed in~\cite{ref:Soko20} to the same end.
(Block error rate refers to the probability that a spatial position---a \textit{block}---contains unrecovered bits after decoding.)
Specifically, the models for both full BP decoding in Sections~\ref{sec:fl_constant_speed}--\ref{sec:fl_pditer} and for sliding window decoding in Section~\ref{sec:sw_lim_iter} already keep track of the number of bits decoded across iterations implicitly; what is required to obtain a scaling law for bit and block error rate is to make use of this knowledge and average the bit and block error rate expressions over the probability distributions of when decoding stops, which we did in~\cite{ref:Soko20} in the context of unlimited number of iterations.

Further, we do not foresee substantial difficulties in extending the scaling laws proposed here to protograph-based ensembles.
What is required to do so is to modify density and mean evolution equations to account for the changed Tanner graph connectivity, as it is done in~\cite{ref:Stin16} for unlimited number of decoding iterations.

\balance



\begin{thebibliography}{10}
\providecommand{\url}[1]{#1}
\csname url@samestyle\endcsname
\providecommand{\newblock}{\relax}
\providecommand{\bibinfo}[2]{#2}
\providecommand{\BIBentrySTDinterwordspacing}{\spaceskip=0pt\relax}
\providecommand{\BIBentryALTinterwordstretchfactor}{4}
\providecommand{\BIBentryALTinterwordspacing}{\spaceskip=\fontdimen2\font plus
\BIBentryALTinterwordstretchfactor\fontdimen3\font minus
  \fontdimen4\font\relax}
\providecommand{\BIBforeignlanguage}[2]{{%
\expandafter\ifx\csname l@#1\endcsname\relax
\typeout{** WARNING: IEEEtran.bst: No hyphenation pattern has been}%
\typeout{** loaded for the language `#1'. Using the pattern for}%
\typeout{** the default language instead.}%
\else
\language=\csname l@#1\endcsname
\fi
#2}}
\providecommand{\BIBdecl}{\relax}
\BIBdecl

\bibitem{ref:Jime99}
A.~{Jimen\'ez Feltstr\"om} and K.~S. {Zigangirov}, ``Time-varying periodic
  convolutional codes with low-density parity-check matrix,'' \emph{IEEE Trans.
  Inf. Theory}, vol.~45, no.~6, pp. 2181--2191, Sep. 1999.

\bibitem{ref:Lent10}
M.~{Lentmaier}, A.~{Sridharan}, D.~J. {Costello}, and K.~S. {Zigangirov},
  ``Iterative decoding threshold analysis for {LDPC} convolutional codes,''
  \emph{IEEE Trans. Inf. Theory}, vol.~56, no.~10, pp. 5274--5289, Oct. 2010.

\bibitem{ref:Kude11}
S.~Kudekar, T.~J. Richardson, and R.~L. Urbanke, ``Threshold saturation via
  spatial coupling: Why convolutional {LDPC} ensembles perform so well over the
  {BEC},'' \emph{IEEE Trans. Inf. Theory}, vol.~57, no.~2, pp. 803--834, Feb.
  2011.

\bibitem{ref:Kude13}
S.~{Kudekar}, T.~{Richardson}, and R.~L. {Urbanke}, ``Spatially coupled
  ensembles universally achieve capacity under belief propagation,'' \emph{IEEE
  Trans. Inf. Theory}, vol.~59, no.~12, pp. 7761--7813, Dec. 2013.

\bibitem{ref:Srid07}
A.~{Sridharan}, D.~{Truhachev}, M.~{Lentmaier}, D.~J. {Costello}, and K.~S.
  {Zigangirov}, ``Distance bounds for an ensemble of {LDPC} convolutional
  codes,'' \emph{IEEE Trans. Inf. Theory}, vol.~53, no.~12, pp. 4537--4555,
  Dec. 2007.

\bibitem{ref:Molo17}
S.~{Moloudi}, M.~{Lentmaier}, and A.~{Graell i Amat}, ``Spatially coupled
  turbo-like codes,'' \emph{IEEE Trans. Inf. Theory}, vol.~63, no.~10, pp.
  6199--6215, Oct. 2017.

\bibitem{ref:Smit12}
B.~P. {Smith}, A.~{Farhood}, A.~{Hunt}, F.~R. {Kschischang}, and J.~{Lodge},
  ``Staircase codes: {FEC} for 100 {Gb}/s {OTN},'' \emph{J. Lightw. Technol.},
  vol.~30, no.~1, pp. 110--117, Jan. 2012.

\bibitem{ref:Aref12}
V.~Aref, N.~Macris, R.~Urbanke, and M.~Vuffray, ``Lossy source coding via
  spatially coupled {LDGM} ensembles,'' in \emph{Proc. IEEE Int. Symp. Inf.
  Theory (ISIT)}, Cambridge, MA, Jul. 2012, pp. 373--377.

\bibitem{ref:Dono13}
D.~L. Donoho, A.~Javanmard, and A.~Montanari, ``Information-theoretically
  optimal compressed sensing via spatial coupling and approximate message
  passing,'' \emph{IEEE Trans. Inf. Theory}, vol.~59, no.~11, pp. 7434--7464,
  Nov. 2013.

\bibitem{ref:Iyen12}
A.~R. {Iyengar}, M.~{Papaleo}, P.~H. {Siegel}, J.~K. {Wolf},
  A.~{Vanelli-Coralli}, and G.~E. {Corazza}, ``Windowed decoding of
  protograph-based {LDPC} convolutional codes over erasure channels,''
  \emph{IEEE Trans. Inf. Theory}, vol.~58, no.~4, pp. 2303--2320, Apr. 2012.

\bibitem{ref:Olmo15}
P.~M. Olmos and R.~L. Urbanke, ``A scaling law to predict the finite-length
  performance of spatially-coupled {LDPC} codes,'' \emph{IEEE Trans. Inf.
  Theory}, vol.~61, no.~6, pp. 3164--3184, Jun. 2015.

\bibitem{ref:Amra09}
A.~Amraoui, A.~Montanari, T.~Richardson, and R.~Urbanke, ``Finite-length
  scaling for iteratively decoded {LDPC} ensembles,'' \emph{IEEE Trans. Inf.
  Theory}, vol.~55, no.~2, pp. 473--498, Feb. 2009.

\bibitem{ref:Stin16}
M.~Stinner and P.~M. Olmos, ``On the waterfall performance of finite-length
  {SC}-{LDPC} codes constructed from protographs,'' \emph{IEEE J. Sel. Areas
  Commun.}, vol.~34, no.~2, pp. 345--361, Feb. 2016.

\bibitem{ref:Cost18}
D.~J. {Costello}, D.~G.~M. {Mitchell}, P.~M. {Olmos}, and M.~{Lentmaier},
  ``Spatially coupled generalized {LDPC} codes: Introduction and overview,'' in
  \emph{Proc. 10th IEEE Int. Symp. Turbo Codes and Iterative Inf. Process.
  (ISTC)}, Hong Kong, China, Dec. 2018.

\bibitem{ref:Soko20}
R.~{Sokolovskii}, A.~{Graell i Amat}, and F.~{Br\"annstr\"om}, ``Finite-length
  scaling of spatially coupled {LDPC} codes under window decoding over the
  {BEC},'' \emph{IEEE Trans. Commun.}, vol.~68, no.~10, pp. 5988--5998, Oct.
  2020.

\bibitem{ref:Kwak21}
H.-Y. Kwak, J.-W. Kim, and J.-S. No, ``Optimizing code parameters of
  finite-length {SC-LDPC} codes using the scaling law,'' \emph{IEEE Access},
  vol.~9, pp. 118\,640--118\,650, Aug. 2021.

\bibitem{ref:Soko21}
R.~{Sokolovskii}, A.~{Graell i Amat}, and F.~{Br\"annstr\"om}, ``On doped
  {SC-LDPC} codes for streaming,'' \emph{IEEE Commun. Lett.}, vol.~25, no.~7,
  pp. 2123--2127, Jul. 2021.

\bibitem{ref:Kwak22}
H.-Y. Kwak, J.-W. Kim, H.~Park, and J.-S. No, ``Optimization of {SC-LDPC} codes
  for window decoding with target window sizes,'' \emph{IEEE Trans. Commun.},
  2022, {Early Access}.

\bibitem{ref:Luby97}
M.~G. Luby, M.~Mitzenmacher, M.~A. Shokrollahi, D.~A. Spielman, and V.~Stemann,
  ``Practical loss-resilient codes,'' in \emph{Proc. Annu. ACM Symp. Theory
  Comput. (STOC)}, El Paso, TX, USA, 1997, pp. 150--159.

\bibitem{ref:Luby01}
M.~G. {Luby}, M.~{Mitzenmacher}, M.~A. {Shokrollahi}, and D.~A. {Spielman},
  ``Efficient erasure correcting codes,'' \emph{IEEE Trans. Inf. Theory},
  vol.~47, no.~2, pp. 569--584, Feb. 2001.

\bibitem{ref:Uhle30}
G.~E. Uhlenbeck and L.~S. Ornstein, ``On the theory of the brownian motion,''
  \emph{Phys. Rev.}, vol.~36, pp. 823--841, Sep. 1930.

\bibitem{ref:Stin16_ppd}
M.~Stinner, L.~Barletta, and P.~M. Olmos, ``Finite-length scaling based on
  belief propagation for spatially coupled {LDPC} codes,'' in \emph{Proc. IEEE
  Int. Symp. Inf. Theory (ISIT)}, Barcelona, Spain, Jul. 2016, pp. 2109--2113.

\bibitem{ref:Pavl14}
G.~A. Pavliotis, \emph{Stochastic Processes and Applications}.\hskip 1em plus
  0.5em minus 0.4em\relax Springer, New York, NY, USA, 2014.

\bibitem{ref:Bene13}
E.~Benedetto, L.~Sacerdote, and C.~Zucca, ``A first passage problem for a
  bivariate diffusion process: Numerical solution with an application to
  neuroscience when the process is {Gauss}-{Markov},'' \emph{J. Comput. Appl.
  Math.}, vol. 242, pp. 41--52, 2013.

\bibitem{ref:FiPy09}
\BIBentryALTinterwordspacing
J.~E. Guyer, D.~Wheeler, and J.~A. Warren, ``{FiPy}: Partial differential
  equations with {P}ython,'' \emph{Comput. Sci. Eng.}, vol.~11, no.~3, pp.
  6--15, 2009. [Online]. Available: \url{http://www.ctcms.nist.gov/fipy}
\BIBentrySTDinterwordspacing

\end{thebibliography}
\end{document}